\documentclass[preprint, aps, prd, superscriptaddress, nofootinbib, floatfix]{revtex4-2}

\usepackage[T1]{fontenc}
\usepackage[utf8]{inputenc}
\usepackage{amsmath, amssymb, physics, bm}
\usepackage{graphicx, xcolor}
\usepackage{siunitx}
\usepackage{ulem}
\usepackage{booktabs}
\usepackage{comment}
\usepackage{stackengine}
\usepackage{adjustbox}
\usepackage{slashed}
\usepackage{multirow}
\usepackage{caption}
\usepackage{geometry,tabularx}
\usepackage{fontawesome}
\usepackage{enumitem}
\usepackage{subcaption}

\newcommand{\ws}{w_s}
\newcommand{\Omr}{\Omega_r}
\newcommand{\Omm}{\Omega_m}
\newcommand{\Omgw}{\Omega_{\mathrm{GW}}}
\newcommand{\fend}{f_{\mathrm{end}}}
\newcommand{\fbeg}{f_{\mathrm{beg}}}
\newcommand{\Hs}{\mathcal{H}}

\newcommand{\Ccal}{\mathcal{C}}
\newcommand{\hprim}{h_{\mathrm{prim}}}
\newcommand{\aentry}{a_{\mathrm{entry}}}
\newcommand{\Tcal}{\mathcal{T}}

\usepackage[colorlinks=true, linkcolor=blue, citecolor=blue, urlcolor=blue]{hyperref}
\usepackage{cleveref}

\begin{document}

\flushbottom

\title{\Large Gravitational Wave Signatures of Cosmological Stasis: A Unified Spectral Template}

\author{Gabriela Barenboim}
\affiliation{Instituto de F\'{i}sica Corpuscular, CSIC-Universitat de Val\`{e}ncia, Paterna 46980, Spain}
\affiliation{Departament de F\'{i}sica Te\`{o}rica, Universitat de Val\`{e}ncia, Burjassot 46100, Spain}
\email{gabriela.barenboim@uv.es}

\author{Anne-Katherine Burns}
\affiliation{Departament de Física Quàntica i Astrofísica (FQA), Universitat de Barcelona (UB), c. Martí i Franqués, 1, 08028 Barcelona, Spain}
\affiliation{Institut de Ciències del Cosmos (ICCUB), Universitat de Barcelona (UB), c. Martí i Franqués, 1, 08028 Barcelona, Spain}
\email{annekatherineburns@icc.ub.edu}

\begin{abstract}

Proposed in 2022 by Dienes et al., stasis is a dynamical fixed point in the early universe in which the equation of state, $w_s$, is fixed at a constant value. In this study we show that the inflationary gravitational wave imprint of any stasis epoch is captured by a closed-form spectral template controlled by two physical inputs, the equation of state $w_s$ and the stasis duration $\Delta N_\mathrm{stasis}$, that applies uniformly
across every microphysical realization. The template presented here yields two independently measurable observables, the spectral tilt of the spectra in the stasis band, $\alpha(w_s)$ and the amplitude step $\mathcal{C}^2(w_s)$ at the beginning and end of the stasis band. Eliminating $w_s$ gives a one-parameter consistency curve $\mathcal{C}^2 = \mathcal{C}^2(\alpha)$ on which the data must lie if the underlying cosmology is any constant-$w$ era. This makes the spectrum falsifiable without knowing $w_s$ in advance: a measured $(\alpha, \mathcal{C}^2)$ pair either lands on the curve or rules out the constant-$w$ class. We show that BBO and DECIGO can resolve the perpendicular displacement from the consistency curve to $\sigma_\perp \simeq 1.5\times10^{-5}$ at a tensor-to-scalar ratio, $r = 0.01$, four orders of magnitude below the curve's range in $\Ccal^2$ meaning that any off-curve deviation is detectable across a broad range of allowed $r$ values.

\end{abstract}

\maketitle

\tableofcontents

\section{Introduction}

The Nobel Prize winning achievement of the first detection of gravitational waves (GWs) by LIGO in 2015 marked the beginning of a new era of scientific discovery. Predicted nearly a century earlier by Einstein in 1916, for many years GWs were thought to be out of reach of any experiment. Today, hundreds of GWs of astrophysical origin have been observed, and many future detectors are being developed to see fainter astrophysical signals as well as the stochastic gravitational-wave background (SGWB), and in particular its primordial component sourced by inflation, which is expected to lie beneath the louder astrophysical foreground. The measurement of the inflationary gravitational wave background (IGWB) would be the first direct observation into the universe at energies above the MeV scale, and could hold clues to further our understanding of its mysterious first fractions of a second. 

The $\Lambda$CDM model of the universe combined with our understanding of inflation gives a clear picture of the expansion history of the universe, the understanding of which is one of the central pursuits of modern cosmology. According to $\Lambda$CDM, after inflation three distinct periods of expansion have occurred, the first driven by radiation, then matter, and today, dark energy. It is generally assumed that in an expanding universe, the relative amounts of these three components, radiation, matter, and dark energy, are constantly evolving and epochs in which there are comparable amounts of any two of these three components are transient. In a series of papers, Dienes et al. have challenged this assumption, introducing cosmological stasis: a dynamical fixed point of the early universe in which the energy fractions $\Omega_i = \rho_i / \rho_\mathrm{tot}$ of multiple components stay exactly constant, even as the universe expands and dilutes~\cite{Dienes:2021woi, Dienes:2022zgd, Dienes:2023ziv, Dienes:2024wnu, Dienes:2025qdw, Dienes:2025tox, Barber:2024vui, Barber:2024izt}. Ordinarily, radiation and matter dilute at different rates ($\rho_r \propto a^{-4}$, $\rho_m \propto a^{-3}$), so their fractions evolve. Stasis is the non-trivial attractor in which this is stopped. In the canonical stasis scenario, $\Omr$ and $\Omm$ are both constant during stasis, and thus the total equation of state is also constant:
\begin{equation}
  \ws = \frac{p}{\rho} = \frac{\Omr}{3} = \mathrm{const.}
  \label{eq:wstasis}
\end{equation}
The original mechanism uses a tower of decaying species~\cite{Dienes:2021woi}, but stasis has since been realized in many microphysical settings: primordial black holes evaporating via Hawking radiation~\cite{Dienes:2022zgd, Dienes:2025qdw}, dynamical scalars~\cite{Dienes:2024wnu}, thermal annihilation~\cite{Barber:2024vui,Barber:2024izt}, field-dependent decay~\cite{Huang:2025odd}, and gravitational interactions~\cite{Long:2025wjw}. All of these realizations share the same fixed-point structure, discussed in detail in Sec.~\ref{sec:stasis}.

A primary observational handle on stasis comes from the stochastic gravitational wave background from inflation. Inflation is known to generate a spectrum of primordial tensor perturbations~\cite{Starobinsky:1979ty, Rubakov:1982df}, which redshift like radiation after re-entering the horizon and contribute to a present-day stochastic background spanning many decades in frequency from the CMB scales probed by Planck and BICEP/Keck down through pulsar timing arrays and into the bands of LISA \cite{LISA:2017pwj}, DECIGO \cite{Kawamura:2011zz}, BBO \cite{Corbin:2005ny}, the Einstein Telescope \cite{Punturo_2010}, and Cosmic Explorer \cite{Reitze:2019iox}. Each mode in this spectrum carries an imprint of the expansion history between its horizon re-entry and the present day, as the amplitude at re-entry depends on the equation of state of the universe at that time. In a purely radiation-dominated cosmology the resulting transfer function gives an essentially flat $h^2\Omega_\mathrm{GW}(f)$ across the relevant bands, but any non-standard expansion phase, i.e. stasis, early matter domination~\cite{Boyle:2005se, Seto:2003kc},  or kination~\cite{Giovannini:1998bp, Giovannini:1999bh, Co:2021lkc} leaves a characteristic, frequency-dependent imprint on the spectrum. The IGWB is therefore a direct probe of the pre-BBN expansion history at energy scales far above those accessible to any other cosmological observable.

During stasis the equation of state, $\ws$, is exactly constant, and as a consequence the expansion during stasis is an exact power law in conformal time, and the tensor mode equation reduces to an exact Bessel equation with no approximations. The spectral imprint can therefore be written in closed form in terms of $\ws$ and the duration $\Delta N_\mathrm{stasis}$, with no residual model dependence on the underlying microphysics. This closed-form structure is what allows the framework developed below: a single template that applies uniformly across all realizations in Table~\ref{tab:realizations}, with predictions that are sharp enough to be falsified by a single measurement.

The remainder of this paper is organized as follows. In Sec.~\ref{sec:stasis} we review stasis as a dynamical fixed point and catalog the microphysical realizations across which our results apply. Sec.~\ref{sec:transfer} derives the inflationary GW transfer function for a generic stasis epoch: because the attractor enforces a constant equation of state $w_s$, the tensor mode equation reduces exactly to a Bessel equation, yielding closed-form expressions for the spectral tilt $\alpha(w_s)$ and the amplitude coefficient $\mathcal{C}^2(\nu)$ at horizon crossing. Sec.~\ref{sec:template} assembles these ingredients into a piecewise spectral template with two free inputs, $w_s$ and the break ratio $f_\mathrm{beg}/f_\mathrm{end}$, and shows how the slope, width, break position, and amplitude step encode the full stasis dynamics; eliminating $w_s$ between $\alpha$ and $\mathcal{C}^2$ gives the consistency curve, which turns the template into a falsifiable prediction. Sec.~\ref{sec:validation} validates the template against the numerical PBH-induced stasis calculation of~\cite{Dienes:2022zgd}. Sec.~\ref{sec:detectability} addresses the degeneracy with constant-$w$ alternatives such as early matter domination, kination, and early dark energy, identifying the consistency relation, the multi-epoch comb, and external particle-physics constraints as the three discriminators. Sec.~\ref{sec:fisher} presents the Fisher forecast quantifying how precisely BBO and DECIGO can measure $(\alpha, \mathcal{C}^2)$ jointly and therefore how sensitively they can test the consistency relation. We conclude in Sec.~\ref{sec:discussion}.

\section{Stasis: definition and realizations}
\label{sec:stasis}

\subsection{Stasis as a dynamical fixed point}

The defining feature of stasis is that the energy fractions $\Omega_i \equiv \rho_i/\rho_\mathrm{tot}$ of multiple cosmological components remain exactly constant even as the universe expands. In ordinary cosmology this is impossible as components with different equations of state $w_i$ redshift at different rates, $\rho_i \propto a^{-3(1+w_i)}$, and thus their fractional contributions to the total energy density evolve by construction. Stasis, however, is a cosmological phenomena realized as a dynamical attractor in a range of beyond-the-Standard-Model settings in which these fractional contributions remain constant over an extended period of time. During stasis source terms in the continuity equations,
\begin{equation}
  \dot\rho_i + 3H(1+w_i)\rho_i = Q_i,
  \label{eq:continuity}
\end{equation}
adjust automatically such that the dilution from Hubble expansion is exactly compensated by the energy injection encoded in $Q_i$, maintaining constant $\Omega_i$ throughout the epoch.

Constant fractions of components with fixed individual equations of state $w_i$ imply that the total equation of state,
\begin{equation}
  \ws = \sum_i w_i \Omega_i,
  \label{eq:ws_def}
\end{equation}
is itself exactly constant during the epoch. The expansion is therefore an exact power law, $a(\tau) \propto \tau^{2/(1+3\ws)}$, with no approximation. As we show in Sec.~\ref{sec:transfer}, this is what makes the GW transfer function during stasis analytically tractable: the tensor mode equation becomes an exact Bessel equation, and all quantities of phenomenological interest can be expressed in closed form in terms of $\ws$ and the duration of the epoch.

\subsection{Microphysical realizations}
\label{sec:realizations}

Stasis was originally introduced in the context of a tower of decaying particles~\cite{Dienes:2021woi}, but has since been realized in a variety of microphysical settings, each governed by its own source-term dynamics in Eq.~\eqref{eq:continuity}.

\paragraph{Decaying tower of species.} The canonical realization uses a tower of particle species with masses $m_j$ and decay widths $\Gamma_j \propto m_j^\gamma$. As successive species decay, they inject energy into the radiation bath at a rate that, for the appropriate spectral index $\gamma$, exactly compensates the dilution of $\Omega_r$ from expansion. The system then locks onto the attractor with $\ws = \Omr/3 \in (0,1/3)$, and the duration of the epoch is controlled by the spectral structure of the tower~\cite{Dienes:2021woi}.

\paragraph{Primordial black holes.} A population of primordial black holes with a continuous mass spectrum evaporating via Hawking radiation can play the role of the decaying tower~\cite{Dienes:2022zgd,Dienes:2025qdw}. The evaporation rate $dM/dt \propto -1/M^2$ together with a power-law mass spectrum gives the right structure to source stasis, with $\ws$ fixed by the spectral index of the mass distribution and the duration set by the ratio of maximum to minimum PBH mass.

\paragraph{Dynamical scalars.} A scalar field rolling in an appropriate potential can also drive stasis~\cite{Dienes:2024wnu}. Unlike the tower realizations, the scalar case can in principle reach $\ws > 1/3$, resulting in an enhancement to the spectra, rather than a suppression, as discussed in more detail in Sec.~\ref{sec:template}.

\paragraph{Thermal annihilation.} Stasis can be sourced by thermal annihilation processes with appropriately chosen $\langle\sigma v\rangle$, in which the annihilation rate sustains the radiation bath against dilution~\cite{Barber:2024vui,Barber:2024izt}.

\paragraph{Field-dependent decay.} Decay rates that depend on a slowly evolving background field provide another mechanism, with the field dynamics controlling the effective $Q_i$~\cite{Huang:2025odd}.

\paragraph{Gravitational interactions.} Most recently, it has been shown that purely gravitational interactions between components can also support a stasis attractor~\cite{Long:2025wjw}, with the coupling strength setting the duration of the epoch.

Table~\ref{tab:realizations} summarizes these mechanisms, giving the range of $\ws$ each can access and the microphysical parameter controlling the duration $\Delta N_\mathrm{stasis}$. The breadth of these microphysical realizations is consistent with the recent argument of~\cite{Halverson:2024oir} that stasis is in fact a generic dynamical outcome rather than a fine-tuned one, arising naturally across broad classes of multi-component cosmologies.

\begin{table}[h]
\centering
\renewcommand{\arraystretch}{1.3}
\begin{tabular*}{\textwidth}{@{\extracolsep{\fill}} l l l l}
\toprule
\textbf{Realization} & $\bm{w_s}$ & \textbf{Duration controlled by} & \textbf{Ref.}\\
\midrule
KK / decaying tower  & $\Omr/3 \in (0,1/3)$ & spectral index of tower & \cite{Dienes:2021woi}\\
PBH-induced          & $\Omr/3$             & PBH mass spectrum       & \cite{Dienes:2022zgd}\\
Dynamical scalar     & potentially $>1/3$   & scalar potential        & \cite{Dienes:2024wnu}\\
Thermal annihilation & model-dependent      & $\langle\sigma v\rangle$ & \cite{Barber:2024vui}\\
Field-dep.\ decay    & model-dependent      & decay function          & \cite{Huang:2025odd}\\
Gravitational        & $\Omr/3$             & coupling strength       & \cite{Long:2025wjw}\\
\bottomrule
\end{tabular*}
\caption{Microphysical realizations of stasis, the equation-of-state parameter $\ws$ each can support, and the microphysical input controlling the duration of the stasis epoch.}
\label{tab:realizations}
\end{table}

\subsection{Universality of the Gravitational Wave imprint}

The microphysics distinguishing these realizations enters the GW spectrum only through two numbers: the equation-of-state parameter $\ws$ during the epoch and its duration $\Delta N_\mathrm{stasis}$. This is due to the fact that the tensor mode equation in an FRW background depends only on the scale factor $a(\tau)$, which is fixed by $\ws$ alone during stasis. The duration $\Delta N_\mathrm{stasis}$ then sets the range of comoving scales that cross the horizon during the epoch. Every other detail of the microphysics, the structure of the tower, the shape of the PBH mass spectrum, the form of the scalar potential, the strength of the gravitational coupling, enters only indirectly, by determining how the underlying mechanism maps onto these two parameters.

For the remainder of this paper we therefore treat $\ws$ and $\Delta N_\mathrm{stasis}$ as the fundamental inputs and derive the GW spectrum in terms of them. Connecting back to a particular realization amounts to specifying which region of the $(\ws, \Delta N_\mathrm{stasis})$ plane that realization populates: the table above is, in effect, a parametric map of where each mechanism sits in this two-dimensional space.

\section{Exact transfer function for a constant-$w_s$ epoch}
\label{sec:transfer}

In this section, we derive the inflationary GW transfer function for a generic stasis epoch in a flat FRW background. We work in conformal time, $\tau$, with primes denoting $d/d\tau$ and define $\Hs = a'/a$. The transverse-traceless tensor perturbation $h_{ij}$, defined via $g_{ij} = a^2(\delta_{ij} + h_{ij})$ with $\partial^i h_{ij} = h^i{}_i = 0$, is decomposed in Fourier modes as
\begin{equation}
  h_{ij}(\mathbf{x},\tau) = \int \frac{d^3k}{(2\pi)^3}
  \sum_{\lambda = +,\times} h^\lambda_{\mathbf{k}}(\tau)\,
  e^\lambda_{ij}(\hat k)\, e^{i\mathbf{k}\cdot\mathbf{x}},
\end{equation}
where $e^\lambda_{ij}(\hat{k})$ are the polarization tensors satisfying $e^\lambda_{ij} e^{\lambda'\, ij} = 2\delta^{\lambda \lambda'}$ and $\hat{k} = \mathbf{k}/|\mathbf{k}|$. Because both polarization states obey the same equation of motion, we drop the polarization label and consider a single mode function $h_k$. The transverse-traceless part of the linearized Einstein equations on the FRW background yields
\begin{equation}
  h_k'' + 2\Hs\,h_k' + k^2 h_k = 0,
\end{equation}
in which the friction term $2\Hs\,h_k'$ is removed by the substitution $\mu_k \equiv a\,h_k$:
\begin{equation}
  \mu_k'' + \left(k^2 - \frac{a''}{a}\right)\mu_k = 0.
  \label{eq:eom}
\end{equation}
Equation~\eqref{eq:eom} is a harmonic oscillator with time-dependent frequency $\omega_k^2(\tau) = k^2 - a''/a$. Sub-horizon modes ($k^2 \gg a''/a$) oscillate as plane waves, while super-horizon modes ($k^2 \ll a''/a$) freeze, with $h_k = \mu_k/a$ approaching a constant.

In any stasis scenario, $\ws$ is by definition exactly constant, fixing the scale factor to the exact power law $a(\tau) \propto \tau^\beta$ with
\begin{equation}
  \beta \equiv \frac{2}{1+3\ws}, \qquad
  \nu \equiv \left|\beta - \tfrac12\right| = \frac{3(1-\ws)}{2(1+3\ws)}.
  \label{eq:beta_nu}
\end{equation}
With $a''/a = \beta(\beta-1)/\tau^2 = (\nu^2 - 1/4)/\tau^2$, Eq.~\eqref{eq:eom} becomes the exact Bessel equation
\begin{equation}
  \mu_k'' + \left[k^2 - \frac{\nu^2 - \tfrac14}{\tau^2}\right]\mu_k = 0,
\end{equation}
with general solution
\begin{equation}
  \mu_k(\tau) = \sqrt{\tau}\bigl[A_k\,J_\nu(k\tau) + B_k\,Y_\nu(k\tau)\bigr].
  \label{eq:bessel-soln}
\end{equation}

Initial conditions are fixed by requiring that on super-horizon scales the mode is frozen at its primordial value, $h_k(\tau) \to \hprim(k)$ as $k\tau \to 0$. Since $Y_\nu(k\tau)$ diverges in this limit, regularity forces $B_k = 0$\footnote{The kination limit $\ws = 1$ corresponds to $\nu = 0$, for which the regularity argument requires a separate treatment: $Y_0(k\tau)$ diverges only logarithmically, and the small-argument expansion of $J_0$ becomes $J_0(k\tau) \to 1$ rather than a power law. The mode-freezing condition $\nu + 1/2 = \beta$ becomes the marginal equality $1/2 = 1/2$, but the conclusion $B_k = 0$ and the form of Eq.~\eqref{eq:WKB} continue to hold.}. Using $J_\nu(k\tau) \approx (k\tau)^\nu / [2^\nu \Gamma(\nu+1)]$ and writing $a(\tau) = A_\mathrm{exp}\,\tau^\beta$, we find
\begin{equation}
  h_k(\tau) = \frac{\mu_k}{a}
  \approx \frac{A_k}{A_\mathrm{exp}} \cdot
  \frac{k^\nu\,\tau^{\nu + 1/2 - \beta}}{2^\nu\,\Gamma(\nu+1)}.
\end{equation}
The identity $\nu + 1/2 = \beta$ makes the $\tau$-dependence drop out exactly, verifying that the mode is frozen on super-horizon scales. Matching this constant to $\hprim(k)$ fixes
\begin{equation}
  A_k = \hprim(k) \cdot \frac{2^\nu\,\Gamma(\nu+1)\,A_\mathrm{exp}}{k^\nu}.
  \label{eq:Ak}
\end{equation}

For $k\tau \gg 1$, the Bessel function takes its asymptotic form
\begin{equation}
  J_\nu(k\tau) \approx \sqrt{\frac{2}{\pi k\tau}}\,\cos\!\left(k\tau - \frac{\nu\pi}{2} - \frac{\pi}{4}\right).
\end{equation}

To express the result in physically transparent variables, we trade $A_\mathrm{exp}$ for the scale factor at horizon entry. Defining horizon entry by $k = \Hs(\tau_\mathrm{entry}) = \beta/\tau_\mathrm{entry}$ gives $\tau_\mathrm{entry} = \beta/k$ and hence
\begin{equation}
  \aentry(k) \equiv a(\tau_\mathrm{entry}) = A_\mathrm{exp}\left(\frac{\beta}{k}\right)^{\!\beta},
  \qquad \text{so} \qquad
  A_\mathrm{exp} = \frac{\aentry(k)\,k^\beta}{\beta^\beta}.
\end{equation}
This convention matches that of Boyle and Steinhardt~\cite{Boyle:2005se}; other sources adopt slightly different choices. Combining with $A_k$ from Eq.~\eqref{eq:Ak} and using $\beta = \nu + 1/2$, the sub-horizon amplitude takes the compact form
\begin{equation}
  h_k(\tau)\big|_\mathrm{WKB}
  = \hprim(k)\,\Ccal(\nu)\,\frac{\aentry(k)}{a(\tau)}\,
    \cos(k\tau - \nu\pi/2 - \pi/4),
  \label{eq:WKB}
\end{equation}
where
\begin{equation}
  \Ccal(\nu) \equiv \frac{2^{\nu+1/2}\,\Gamma(\nu+1)}{\sqrt{\pi}\,\beta^\beta}
  \label{eq:Ccal}
\end{equation}
is the Bessel amplitude coefficient. Because $\Ccal(\nu)$ is $k$-independent, it rescales the overall transfer function without affecting its spectral shape. Table~\ref{tab:C_val} collects its value for several benchmark equations of state.

\begin{table}[h]
  \centering
  \renewcommand{\arraystretch}{1.3}
  \setlength{\tabcolsep}{12pt}
  \begin{tabular}{ccccc}
    \toprule
    $\ws$ & $\beta$ & $\nu$ & $\Ccal(\nu)$ & $\Ccal^2(\nu)$ \\
    \midrule
    $1/3$ (RD)     & $1$   & $1/2$  & $1$              & $1$              \\
    $0$ (MD)       & $2$   & $3/2$  & $3/4$            & $9/16$           \\
    $1/2$          & $4/5$ & $3/10$ & $\approx1.054$  & $\approx1.11$   \\
    $1$ (kination) & $1/2$ & $0$    & $\approx1.13$   & $\approx1.27$   \\
    \bottomrule
  \end{tabular}
  \caption{The exponent $\beta$, Bessel order $\nu$, and amplitude coefficient $\Ccal(\nu)$ of Eq.~\eqref{eq:Ccal} evaluated for several representative equations of state. The squared coefficient $\Ccal^2(\nu)$ sets the modification to the GW energy density relative to radiation domination, which is recovered as $\Ccal = 1$ at $\ws = 1/3$.}
  \label{tab:C_val}
\end{table}

\subsection{From the WKB amplitude to $\Omega_\mathrm{GW}$}

The present-day energy density of the inflationary GW background is conventionally written as
\begin{equation}
  h^2\Omgw(k,\tau_0)
  = \frac{1}{12}\left(\frac{k}{a_0 H_0}\right)^{\!2}
    \mathcal{P}_T(k)\,\overline{|\Tcal(k)|^2},
  \label{eq:Omega_std}
\end{equation}
where $\mathcal{P}_T(k) = A_T(k/k_*)^{n_T}$ is the primordial tensor power spectrum and $\Tcal(k) \equiv h_k(\tau_0)/\hprim(k)$ is the GW transfer function, with the overline denoting an average over the oscillation phase. Using Eq.~\eqref{eq:WKB},
\begin{equation}
  \overline{|\Tcal(k)|^2}
  = \frac{\Ccal^2(\nu)}{2}\left(\frac{\aentry(k)}{a_0}\right)^{\!2},
  \label{eq:Tcal_squared}
\end{equation}
so the $k$-dependence of $\Omgw$ reduces to that of $\aentry(k)$. During an era with equation of state $\ws$, $H \propto a^{-3(1+\ws)/2}$, and the horizon-entry condition $k = aH$ gives
\begin{equation}
  \aentry(k) \propto k^{-2/(1+3\ws)} = k^{-\beta}.
  \label{eq:aentry_k}
\end{equation}
Combining Eqs.~\eqref{eq:Omega_std}--\eqref{eq:aentry_k} yields
\begin{equation}
  h^2\Omgw(k) \propto k^{2} \cdot k^{n_T} \cdot k^{-2\beta}
  = k^{\,n_T + 2 - 4/(1+3\ws)},
  \label{eq:Omgw_spectral}
\end{equation}
where the three factors come from the prefactor $(k/a_0H_0)^2$, the primordial tilt $\mathcal{P}_T \propto k^{n_T}$, and the transfer function $\aentry^2 \propto k^{-2\beta}$ respectively.

Setting $\ws = 1/3$ in Eq.~\eqref{eq:Omgw_spectral} recovers the standard radiation-era result $\Omgw \propto k^{n_T}$. For generic $\ws$, the modified $\aentry(k)$ scaling shifts the spectral index, leaving a spectral distortion
\begin{equation}
  \alpha(\ws) \equiv n_\mathrm{stasis} - n_\mathrm{RD}
  = \frac{2(3\ws - 1)}{1+3\ws}
  \label{eq:alpha}
\end{equation}
relative to the radiation-dominated baseline. The sign and magnitude of $\alpha$ encode the qualitative effect on the GW spectrum:
\begin{itemize}
  \item For $\ws < 1/3$, $\alpha < 0$: a red tilt and suppression of the spectrum. The matter-dominated limit $\ws = 0$ gives $\alpha = -2$, reproducing the well-known $\Omgw \propto f^{n_T-2}$ suppression of an early matter era.
  \item For $\ws > 1/3$, $\alpha > 0$: a blue tilt and enhancement, as in dynamical-scalar stasis. The kination limit $\ws = 1$ saturates at $\alpha = +1$.
  \item For $\ws = 1/3$, $\alpha = 0$ and the epoch is spectrally indistinguishable from radiation domination.
\end{itemize}

We emphasize that the transfer-function algebra above is not new. The reduction of the tensor mode equation to a Bessel equation for a constant-$w$ background, and the matching of frozen super-horizon modes to the WKB sub-horizon regime, is the construction of Boyle and Steinhardt~\cite{Boyle:2005se}, building on the earlier analysis of tensor spectra across cosmological transitions by Giovannini~\cite{Giovannini:1998bp, Giovannini:1999bh}. Here, we show that because $\ws$ is exactly constant throughout a stasis epoch, the power law $a(\tau)\propto\tau^\beta$ is not an approximation but a defining property, and the Bessel solution \eqref{eq:bessel-soln} is correspondingly exact. This lets us read the same algebra in stasis-native variables: the equation of state $\ws$ and the duration $\Delta N_\mathrm{stasis}$ become the two physical inputs, and the spectral distortion \eqref{eq:alpha} follows as a closed-form function of $\ws$ alone. The payoff is the consistency relation, discussed in detail in Sec.~\ref{sec:detectability}: a single curve, given by Eq.~\eqref{eq:consistency_curve}, on which every constant-$w$ era must lie, turning a known transfer function into a falsifiable prediction.

\section{Piecewise spectral template and the consistency relation}
\label{sec:template}

The full gravitational wave spectrum is set by the time at which each mode crossed the cosmological horizon. Higher-frequency modes entered the horizon earlier, while lower-frequency modes entered later. In stasis scenarios, this naturally partitions the spectrum into three physically distinct bands. The low-frequency band, $f < \fend$, consists of modes that crossed the horizon after stasis ends, during the subsequent standard radiation-dominated (RD) era. These modes never experienced the stasis epoch as super-horizon perturbations and inherit the unmodified RD template. The intermediate stasis band, $\fend < f < \fbeg$, contains modes that entered the horizon during stasis itself, so their initial amplitudes are set by the stasis-era Bessel transfer function and their subsequent evolution carries the imprint of the modified equation of state. The high-frequency band, $f > \fbeg$, comprises modes that crossed the horizon before stasis began, in the earlier pre-stasis RD era. Although they entered as ordinary RD modes, they were already sub-horizon when stasis commenced and were therefore subject to its diluting and tilting effects throughout that epoch. The slopes and amplitude factors of these three bands, expressed relative to the standard RD template, are summarized below:

\begin{table}[h]
\renewcommand{\arraystretch}{1.35}
\setlength{\tabcolsep}{12pt}
\begin{center}
\begin{tabular}{clll}
\toprule
\textbf{Band} & \textbf{Horizon entry} & \textbf{Slope} & \textbf{Amplitude factor}\\
\midrule
$f < \fend$         & post-stasis RD & $n_T$               & $1$\\
$\fend < f < \fbeg$ & during stasis  & $n_T + \alpha(\ws)$ & $\Ccal^2(\nu)$\\
$f > \fbeg$         & pre-stasis RD  & $n_T$               & $(\fbeg/\fend)^{\alpha}$\\
\bottomrule
\end{tabular}
\caption{The three frequency bands of the gravitational wave spectrum, classified by when their modes crossed the cosmological horizon. Slopes and amplitude factors are given relative to the standard radiation-dominated (RD) template.}
\end{center}
\end{table}

Note that $\Ccal^2(\nu)$ only appears in the stasis band. Modes with $f > \fbeg$ crossed the horizon before stasis began, entirely in the pre-stasis RD era, so their Bessel coefficient is $\Ccal(\nu_\mathrm{RD}) = 1$. These modes are affected by the factor $(\fbeg/\fend)^{\alpha}$ due to the fact that their $\Omega_{GW}$ fraction drifted during the stasis epoch, even though they were already sub-horizon, as described in more detail in Sec.~\ref{sec:accumulated_tilt}.

The following is the full template. Here $h^2\Omgw^{(\mathrm{RD})}$ is the standard IGWB in a purely RD universe, including its $g_*$-induced fine structure.
\begin{equation}
h^2\Omgw(f)
= h^2\Omgw^{(\mathrm{RD})}(f) \times
\begin{cases}
  1
  & f < \fend\\[8pt]
  \Ccal^2(\nu)
  \cdot\!\left(\dfrac{f}{\fend}\right)^{\!\alpha(\ws)}
  & \fend < f < \fbeg\\[10pt]
  \left(\dfrac{\fbeg}{\fend}\right)^{\!\alpha(\ws)}
  & f > \fbeg
\end{cases}
\label{eq:template}
\end{equation}
in which $\alpha(\ws) = 2(3\ws-1)/(1+3\ws)$ and $\Ccal^2(\nu)$ is given by Eq.~\eqref{eq:Ccal}.

\subsection{The accumulated tilt of sub-horizon modes}
\label{sec:accumulated_tilt}

A key feature of the stasis-modified spectrum is that the gravitational wave energy fraction $\Omgw(k)$ drifts during stasis even for modes that were already sub-horizon when the epoch began in which $f > \fbeg$. The mechanism is straightforward: gravitational waves always redshift like radiation once inside the horizon, so $\rho_{GW}(k) \propto a^{-4}$ at all times, but the critical density during stasis follows the modified equation of state, $\rho_c \propto a^{-3(1+\ws)}$. Their ratio therefore evolves as
\begin{equation}
\Omgw(k,\tau)\big|_{\text{stasis}} \propto \frac{a^{-4}}{a^{-3(1+\ws)}} = a^{3\ws - 1},
\end{equation}
which is constant only in the pure radiation-dominated case $\ws = 1/3$. For any $\ws \neq 1/3$, every sub-horizon mode's energy fraction drifts throughout the stasis epoch, regardless of when it originally crossed the horizon. Integrating from the onset of stasis at $a_\mathrm{beg}$ to its end at $a_\mathrm{end}$ gives
\begin{equation}
\frac{\Omgw(k,a_\mathrm{end})}{\Omgw(k,a_\mathrm{beg})} = \left(\frac{a_\mathrm{end}}{a_\mathrm{beg}}\right)^{3\ws-1} = e^{(3\ws-1)\Delta N}.
\end{equation}
The duration of stasis $\Delta N$ is fixed by the frequency ratio $\fbeg/\fend$ through Eq.~\eqref{eq:width}, which yields $(3\ws-1)\Delta N = \frac{2(3\ws-1)}{1+3\ws} \ln(\fbeg/\fend) = \alpha(\ws)\ln(\fbeg/\fend)$ and therefore
\begin{equation}
\frac{\Omgw(k,a_\mathrm{end})}{\Omgw(k,a_\mathrm{beg})} = \left(\frac{\fbeg}{\fend}\right)^{\!\alpha(\ws)}.
\label{eq:accumulated_tilt}
\end{equation}

This factor applies to all modes that were sub-horizon during stasis, including those in the pre-stasis band $f > \fbeg$ that crossed the horizon long before the epoch began. It is tempting to treat the accumulated tilt as physically distinct from the entropy growth that occurs over the same interval, the former arising from the scaling of $\rho_c$, the latter from energy injection into the radiation bath, and to include both as independent multiplicative factors. However, this would be a double-counting. Both effects are consequences of the same underlying fact, namely that $\rho_c \propto a^{-3(1+\ws)}$ rather than $a^{-4}$ during stasis. The accumulated tilt and the entropy dilution are therefore two descriptions of the same dynamics, not two independent effects.

A short calculation makes this explicit: $T \propto \rho_r^{1/4} \propto a^{-3(1+\ws)/4}$ gives $S_r \propto T^3 a^3 \propto a^{3(1-3\ws)/4}$, so $\Delta = (a_\mathrm{end}/a_\mathrm{beg})^{3(1-3\ws)/4}$ and
\begin{equation}
\Delta^{-4/3} = \left(\frac{a_\mathrm{end}}{a_\mathrm{beg}}\right)^{\!3\ws-1} = \left(\frac{\fbeg}{\fend}\right)^{\!\alpha(\ws)}.
\label{eq:delta_equals_tilt}
\end{equation}

Here, we retain $(\fbeg/\fend)^{\alpha(\ws)}$, which follows directly from tracking $\Omgw$ through the three phases and makes the spectral structure manifest. This is a feature of stasis specifically: because energy transfer from the decaying component into radiation occurs continuously throughout the epoch at exactly the rate required to keep $\Omega_r$ constant, the entropy growth and the drift of $\Omgw$ are inseparable. The situation differs from standard non-stasis non-RD eras such as early matter domination, where the matter component redshifts as dust through most of the era and decays into radiation in a relatively brief window at the end when $\Gamma \sim H$; there the entropy production is concentrated at the end and is conventionally treated as a separate step from the in-era drift of $\Omgw$. In stasis no such separation exists, and the template must be written accordingly.

\subsection{Physical structure of the two breaks}
\label{sec:breaks}

With the accumulated tilt included, the breaks have a clean physical interpretation. Both are characterized by the same ratio $\Ccal^2(\nu)$, reflecting the fact that the Bessel mismatch between stasis-era and RD-era horizon crossing is the only new physics introduced at each transition. In this work we model these breaks as a sharp transition in the spectra. In our forthcoming work, we will model the smooth physical transition from RD to the stasis epoch and back to RD \cite{BarenboimBurns:stasis3}. 

\paragraph{Lower break at $\fend$.} Crossing from the post-stasis RD band into the stasis band, the amplitude jumps by $\Ccal^2(\nu)$ and the slope shifts from $n_T$ to $n_T + \alpha(\ws)$:
\begin{equation}
\frac{h^2\Omgw(\fend^+)}{h^2\Omgw(\fend^-)} = \Ccal^2(\nu).
\label{eq:lower_jump}
\end{equation}
The jump is set entirely at horizon crossing: modes just above $\fend$ entered the horizon during stasis and acquire the stasis-era Bessel coefficient, while modes just below entered after stasis ended and carry the standard RD value $\Ccal(\nu_\mathrm{RD}) = 1$. For canonical stasis $\Ccal^2 \leq 1$, so the spectrum drops at $\fend$.

\paragraph{Upper break at $\fbeg$.} At the upper break, from below and above respectively:
\begin{align}
h^2\Omgw(\fbeg^-) &= h^2\Omgw^{(\mathrm{RD})}(\fbeg)\cdot \Ccal^2(\nu)\cdot \!\left(\frac{\fbeg}{\fend}\right)^{\!\alpha},\\
h^2\Omgw(\fbeg^+) &= h^2\Omgw^{(\mathrm{RD})}(\fbeg)\cdot \!\left(\frac{\fbeg}{\fend}\right)^{\!\alpha}.
\end{align}
The accumulated tilt factor is common to both sides and cancels in the ratio:
\begin{equation}
\frac{h^2\Omgw(\fbeg^-)}{h^2\Omgw(\fbeg^+)} = \Ccal^2(\nu).
\label{eq:upper_jump}
\end{equation}
The discontinuity at the upper break is again purely the Bessel mismatch between modes that crossed the horizon in stasis versus in RD. The symmetry between the two breaks is a useful consistency check: a measured spectrum exhibiting markedly different jump ratios at $\fend$ and $\fbeg$ would be inconsistent with a single stasis epoch.

\paragraph{Ratio of the two flat plateaus.} The post-stasis plateau ($f < \fend$) has amplitude factor $1$; the pre-stasis plateau ($f > \fbeg$) has factor $(\fbeg/\fend)^\alpha$. Their ratio is therefore
\begin{equation}
\frac{h^2\Omgw(f > \fbeg)}{h^2\Omgw(f < \fend)} = \left(\frac{\fbeg}{\fend}\right)^{\!\alpha(\ws)}.
\label{eq:plateau_ratio}
\end{equation}
The plateau ratio is a direct measure of the accumulated tilt and depends on only two quantities: the equation-of-state parameter $\ws$ (through $\alpha$) and the duration of stasis (through $\fbeg/\fend$). Both are independently accessible from other spectral features, so the plateau ratio provides a redundant consistency check on the stasis interpretation rather than an independent parameter.

\subsection{Width of the feature}

During stasis, $aH \propto a^{-(1+3\ws)/2}$, so
\begin{equation}
  \ln\frac{\fbeg}{\fend} = \frac{1+3\ws}{2}\,\Delta N_\mathrm{stasis}.
  \label{eq:width}
\end{equation}
For fixed $\Delta N$, wider log-frequency features correspond to larger $\ws$.

\subsection{Break frequencies in physical units}

The lower break is set by the Hubble scale when stasis ends (post-decay temperature $T_\mathrm{end}$):
\begin{equation}
  \fend \approx 2.63\times10^{-8}\,\mathrm{Hz}
  \left(\frac{T_\mathrm{end}}{1\,\mathrm{GeV}}\right)
  \left(\frac{g_*(T_\mathrm{end})}{106.75}\right)^{1/2}
  \!\!\left(\frac{g_{*S}(T_\mathrm{end})}{106.75}\right)^{-1/3}.
  \label{eq:fend}
\end{equation}
The upper break follows from Eq.~\eqref{eq:width}:
\begin{equation}
  \fbeg = \fend \cdot
  \exp\!\left(\frac{1+3\ws}{2}\,\Delta N_\mathrm{stasis}\right)
  \cdot\left(\frac{g_*(T_\mathrm{beg})}{g_*(T_\mathrm{end})}\right)^{1/2}
  \!\!\left(\frac{g_{*S}(T_\mathrm{end})}{g_{*S}(T_\mathrm{beg})}\right)^{1/3}.
  \label{eq:fbeg}
\end{equation}
The $g_*$ ratio is $\mathcal{O}(1)$ unless stasis straddles the QCD or electroweak transitions.

\subsection{Entropy production as a derived observable}
\label{sec:entropy_derived}

The identification $\Delta^{-4/3} = (\fbeg/\fend)^\alpha$ established in Eq.~\eqref{eq:delta_equals_tilt} can be inverted to give the total entropy produced during stasis as a function of spectral observables alone:
\begin{equation}
  \Delta = \left(\frac{\fbeg}{\fend}\right)^{\!-3\alpha/4}.
  \label{eq:Delta_from_observables}
\end{equation}
Although $\Delta$ does not enter the GW template as a free parameter this formula is still a meaningful result. It gives the entropy budget of any stasis epoch directly from two quantities that are read off the spectrum: the spectral distortion $\alpha$ from the stasis-band slope and the break frequency ratio $\fbeg/\fend$ from the feature width.

The derivation uses only two ingredients: the defining property of stasis, $\Omr = \mathrm{const}$, and the approximation that $g_*$ is roughly constant during the epoch. No property of the energy-injecting species enters. Eq.~\eqref{eq:Delta_from_observables} therefore holds for any stasis realization satisfying these two conditions, irrespective of whether the underlying mechanism is a decaying tower, PBH evaporation, a dynamical scalar, gravitational interactions, or thermal annihilation. In this sense it plays the role of an equation-of-state relation: knowing $(\ws, \Delta N_\mathrm{stasis})$ is sufficient to determine the total entropy production, without reference to the details of the interactions that produced it.

Practically, Eq.~\eqref{eq:Delta_from_observables} provides a cross-check between the GW signature of a candidate stasis epoch and the entropy budget predicted by its microscopic realization. Given values of $\ws$ and $\Delta N$ inferred from the spectrum, the entropy produced during stasis can be compared against the entropy released by, e.g., PBH evaporation~\cite{Dienes:2022zgd,Dienes:2025qdw} or by the decaying tower~\cite{Dienes:2021woi}, providing an independent consistency test of the realization.

\subsection{The consistency relation}
\label{sec:consistency}

The template of Eq.~\eqref{eq:template} yields two independently measurable observables from the GW spectrum: the spectral tilt $\alpha$ in the stasis band and the amplitude step $\Ccal^2$ at either break. Both are determined by the single number $\ws$ through Eqs.~\eqref{eq:alpha} and~\eqref{eq:Ccal}. Eliminating $\ws$ between these two relations gives a one-dimensional curve in the $(\alpha, \Ccal^2)$ plane,
\begin{equation}
\Ccal^2 = \Ccal^2\!\left(\alpha\right) \equiv \left[\frac{2^{\nu(\alpha)+1/2}\,\Gamma(\nu(\alpha)+1)}{\sqrt{\pi}\,\beta(\alpha)^{\beta(\alpha)}}\right]^2, \quad \nu = \frac{1-\alpha}{2}, \quad \beta = \frac{2-\alpha}{2}.
\label{eq:consistency_curve}
\end{equation}
This is the consistency relation: any constant-$w$ era must produce a $(\alpha, \Ccal^2)$ pair that lies on this curve. Measuring $\alpha$ from the spectral slope completely determines the predicted $\Ccal^2$, which can then be compared independently against the measured amplitude step at $\fend$. A confirmed match is a non-trivial internal test of the constant-$w$ hypothesis; any discrepancy falsifies it. We discuss in detail how this discriminates stasis from its mimics in Sec.~\ref{sec:detectability}, and quantify the precision with which BBO and DECIGO can perform this test in Sec.~\ref{sec:fisher}.

\subsection{How to read the spectrum}

\begin{figure}
    \centering
    \includegraphics[width=0.9\linewidth]{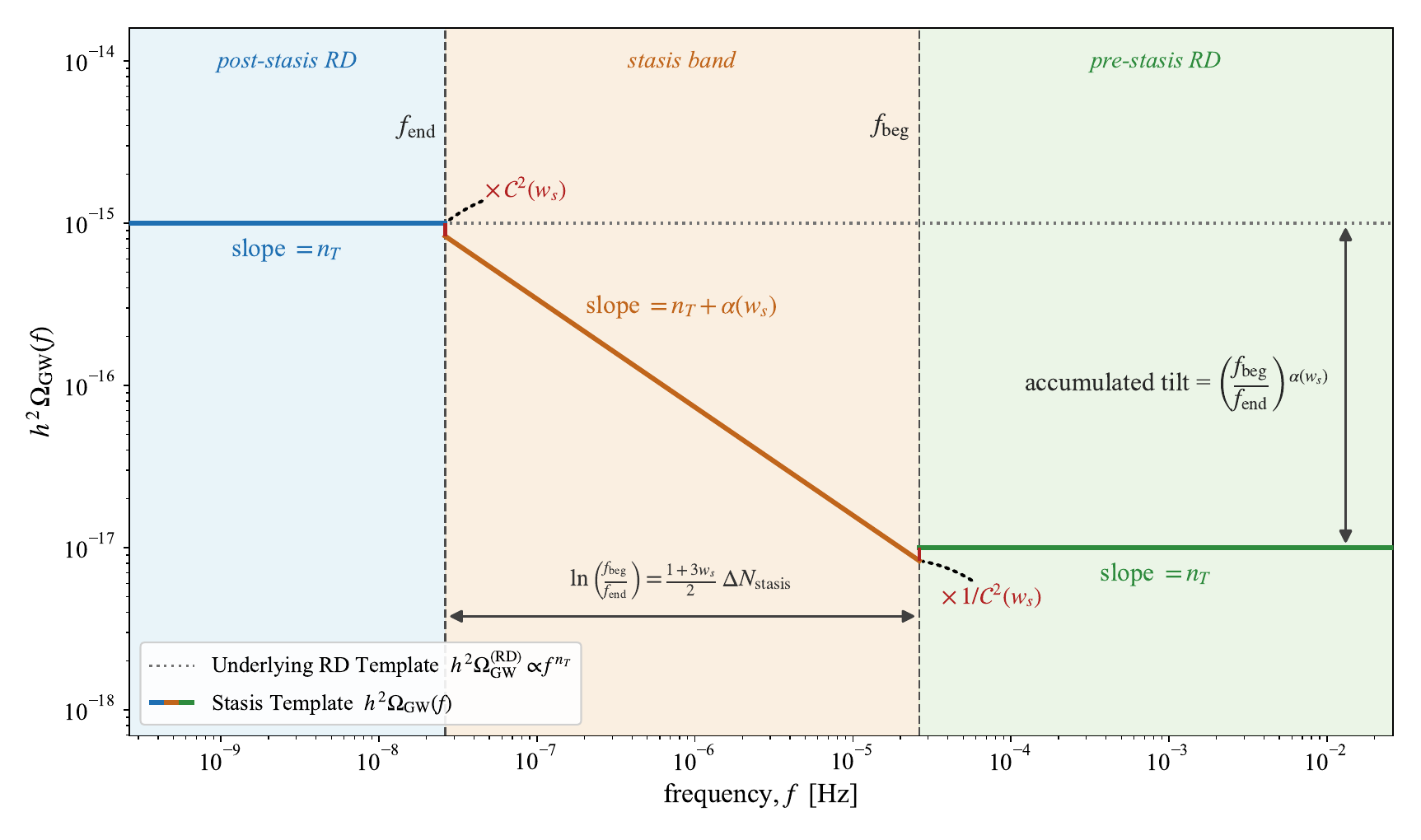}
    \caption{Piecewise gravitational-wave spectral template of Eq.~\eqref{eq:template} for a single stasis epoch, shown with canonical, illustrative parameters. The spectrum splits into three bands set by horizon-crossing time: the post-stasis RD plateau ($f<\fend$), the stasis band ($\fend<f<\fbeg$) with tilted slope $n_T+\alpha(\ws)$, and the pre-stasis RD plateau ($f>\fbeg$). The dotted grey line is the unmodified RD template $h^2\Omgw^{(\mathrm{RD})}\propto f^{\,n_T}$. We choose to set $n_T=0$ and $\ws=1/6$ with arbitrary break frequencies as an example. Both breaks carry the same Bessel mismatch coefficient, $\Ccal^2(\nu)=0.830$. The pre-stasis plateau sits below the post-stasis one by the accumulated-tilt factor $(\fbeg/\fend)^{\alpha}=0.01$.}
    \label{fig:stasis_anatomy}
\end{figure}

Each spectral feature in Eq.~\eqref{eq:template} maps onto a specific physical quantity, so a measurement of the full feature determines the underlying stasis dynamics without degeneracy. The slope of the stasis band, $n_T + \alpha(\ws)$, fixes the equation-of-state parameter $\ws$ through $\alpha(\ws) = 2(3\ws-1)/(1+3\ws)$. The logarithmic width of the feature, $\ln(\fbeg/\fend)$, then fixes the duration of stasis via Eq.~\eqref{eq:width}: $\Delta N_\mathrm{stasis} = (2/(1+3\ws))\ln(\fbeg/\fend)$. The position of the lower break $\fend$ fixes the post-stasis temperature $T_\mathrm{end}$ via Eq.~\eqref{eq:fend}, and hence the cosmological epoch in which stasis terminated. Finally, the amplitude step at either break provides a redundant determination of $\Ccal^2(\nu)$, which is itself fixed by $\ws$; consistency between the measured jump and the value predicted from the slope via Eq.~\eqref{eq:consistency_curve} is a non-trivial test of the single-epoch stasis interpretation. A measured spectrum that fits the stasis band's slope and width but shows an inconsistent jump ratio at either break would point to either a more complex epoch structure or to physics outside the template. Figure~\ref{fig:stasis_anatomy} shows each of these spectral features using canonical illustrative parameter values.

\section{Validation against PBH-induced stasis}
\label{sec:validation}

To test the template against an independent calculation, we compare its predictions to the numerical PBH-induced stasis spectrum of~\cite{Dienes:2022zgd}. In this realization stasis is sourced by a population of primordial black holes with a power-law mass spectrum $dn/dM \propto M^{\alpha_\mathrm{PBH}}$ between $M_\mathrm{min}$ and $M_\mathrm{max}$, evaporating via Hawking radiation into the radiation bath. The stasis equation of state and duration are not free parameters in this setup, but are fixed by the spectral index of the PBH mass distribution through the analytic relations~\cite{Dienes:2022zgd},
\begin{equation}
  \ws = -\frac{\alpha_\mathrm{PBH} + 1}{\alpha_\mathrm{PBH} + 7},
  \qquad
  \Omega_\mathrm{BH} = \frac{4\alpha_\mathrm{PBH} + 10}{\alpha_\mathrm{PBH} + 7},
  \qquad
  \Delta N_\mathrm{stasis} \approx \frac{\alpha_\mathrm{PBH} + 7}{3}\ln\!\frac{M_\mathrm{max}}{M_\mathrm{min}},
  \label{eq:PBH_relations}
\end{equation}
which we use directly. Note that $\alpha_\mathrm{PBH}$ here denotes the spectral index of the PBH mass distribution and is not to be confused with the stasis spectral distortion $\alpha(\ws)$ of Eq.~\eqref{eq:alpha}.

\subsection{Comparison with the benchmark of Fig.~6 of~\cite{Dienes:2022zgd}}

\begin{figure}
    \centering
    \includegraphics[width=0.9\linewidth]{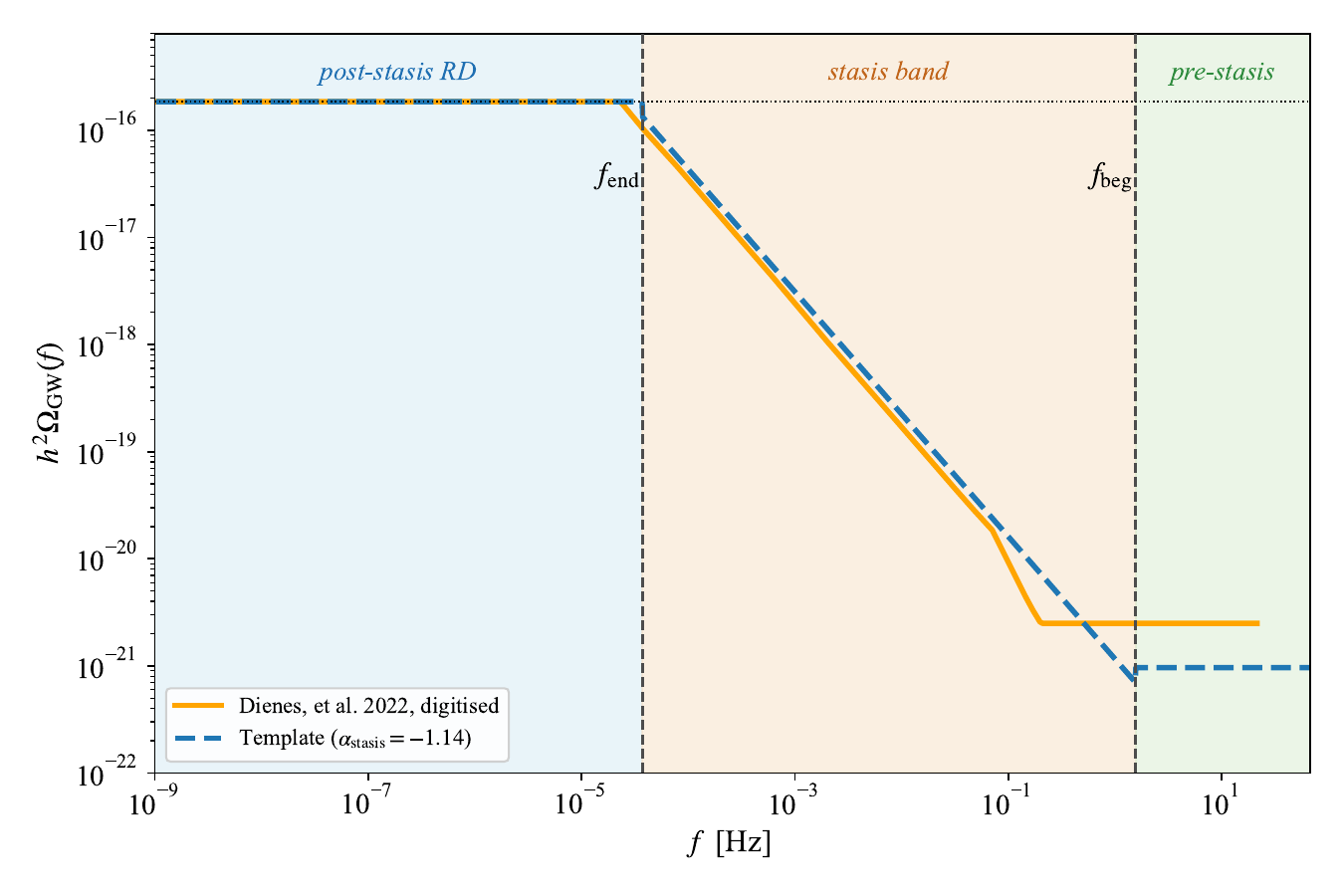}
    \caption{Comparison between the analytic template of Eq.~\eqref{eq:template} (blue dashed line) and the digitized orange curve from Fig.~6 (left panel) of~\cite{Dienes:2022zgd} (orange line), corresponding to PBH-induced stasis with $\alpha_\mathrm{PBH} = -3/2$, $M_\mathrm{min} = 100$\,g, $M_\mathrm{max} = 10^5$\,g. The post-stasis plateau, the amplitude jump at $\fend$, and the slope of the stasis band are template predictions with no free parameters. The deviation at high frequencies is expected: see the discussion of pre-stasis cosmology in the text.}
    \label{fig:validation_orange}
\end{figure}

We digitized the orange curve from the left panel of Fig.~6 of~\cite{Dienes:2022zgd}, corresponding to the benchmark $\alpha_\mathrm{PBH} = -3/2$, $M_\mathrm{min} = 100$\,g, $M_\mathrm{max} = 10^5$\,g. Through Eq.~\eqref{eq:PBH_relations} this fixes $\ws \approx 0.091$, $\alpha(\ws) \approx -1.14$, and $\Ccal^2(\nu) \approx 0.72$. The break frequencies $\fend$ and $\fbeg$ and the post-stasis plateau amplitude are read directly from the digitized data, avoiding any modeling uncertainty in the temperature-frequency map of Eq.~\eqref{eq:fend}. We then overlay the analytic template of Eq.~\eqref{eq:template} with no free parameters. Fig.~\ref{fig:validation_orange} shows the result.

Three features of the template match the numerical calculation. The post-stasis plateau at $f < \fend$ sits at the standard RD value, the point at which the spectra starts dropping begins at the predicted $\fend$, and the slope of the stasis band tracks $\alpha(\ws) \approx -1.14$ across roughly five decades in frequency. The predicted amplitude suppression $\Ccal^2(\nu) \approx 0.72$ at the break is a $0.14$ decade effect, comparable to the digitization precision here, and so is consistent with this comparison and is absorbed into the overall normalization agreement across the band.

Above $f \sim 0.3$\,Hz the digitized curve flattens out at $h^2\Omgw \sim 2 \times 10^{-21}$, while the template continues to descend before flattening at a higher $\fbeg \sim 4$\,Hz. This deviation is expected. The Dienes, et al. spectrum reflects the full pre-stasis cosmology of the PBH realization, which is not standard radiation domination. From Eq.~(4.17) of~\cite{Dienes:2022zgd}, the spectrum above $\fend$ comprises four segments rather than two: (i) the stasis band with slope $\alpha(\ws)$; (ii) a PBH-domination band with slope $\xi(0) = -2$; (iii) a PBH-formation band with slope $\xi(w_c)$; and (iv) an inflationary cutoff. The apparent ``upper break'' at $f \sim 0.3$\,Hz visible in the data is in fact the boundary between the PBH-domination band and the PBH-formation band, not the end of stasis itself. The true end-of-stasis break sits at lower frequency, where the digitized data first transitions from slope $\alpha(\ws) \approx -1.14$ to slope $-2$.

The validation against the PBH benchmark therefore tests the predictions intrinsic to the stasis epoch with no free parameters. With the break frequencies and the post-stasis plateau amplitude anchored to the digitized data, the genuine zero-parameter prediction is the stasis-band slope $\alpha(\ws) \approx -1.14$, which the comparison confirms across roughly five decades in frequency; the accompanying amplitude factor $\Ccal^2(\nu) \approx 0.72$ is consistent with the data but, as a $0.14$-decade effect, is not independently resolved here. The high-frequency continuation above $\fbeg$ is governed by the cosmology preceding stasis rather than by the stasis epoch itself: the template assumes a standard RD pre-history, whereas this benchmark is fixed by the pre-stasis PBH cosmology, so the deviation there is expected and does not bear on the stasis-band predictions.

\subsection{What this validation does and does not establish}

One quantitative prediction of the template cannot be tested against the PBH realization at all. The plateau-ratio prediction of Eq.~\eqref{eq:plateau_ratio},
\begin{equation}
  \frac{h^2\Omgw(f > \fbeg)}{h^2\Omgw(f < \fend)}
  = \left(\frac{\fbeg}{\fend}\right)^{\!\alpha(\ws)},
\end{equation}
assumes a standard radiation-dominated era immediately preceding stasis, so that the spectrum at $f > \fbeg$ is a flat plateau set by the accumulated drift of $\Omgw$ during the stasis epoch. In PBH-induced stasis this assumption fails: the four-segment structure of~\cite{Dienes:2022zgd} replaces the flat pre-stasis plateau with a non-trivial multi-band continuation, and the single-number prediction $(\fbeg/\fend)^\alpha$ does not have a counterpart in the data. Testing the plateau-ratio prediction quantitatively therefore requires a stasis realization for which the pre-stasis cosmology is itself standard RD, the KK decaying-tower scenario of~\cite{Dienes:2021woi} or the dynamical-scalar realization of~\cite{Dienes:2024wnu} are the natural candidates, and we defer this to future work.

Within the regime it can address, the comparison confirms the prediction that depends only on the stasis epoch itself and is not used to anchor the template: the stasis-band slope $\alpha(\ws) \approx -1.14$, recovered across roughly five decades in frequency. The accompanying amplitude factor $\Ccal^2(\nu) \approx 0.72$ is consistent with the data but, as a $0.14$-decade effect, is not independently resolved at this benchmark, while $\fend$ and $\fbeg$ are anchored to the digitized data rather than predicted. Therefore, we have been able to show that the template reproduces the stasis-band behavior in this realization, while the pre-stasis structure of the spectrum is realization-dependent and must be supplied separately. 

\subsection{Robustness against mergers and accretion}
\label{sec:mergers_accretion}

A subsequent analysis of PBH stasis~\cite{Dienes:2025qdw} examines how the mass distribution evolves under PBH mergers and accretion, raising the question of whether these effects modify the stasis parameters $(\ws, \Delta N)$ underlying the validation above. The conclusion is that they do not, in the regime relevant here.

Mergers are negligible throughout the parameter space of phenomenological interest, leaving the PBH mass spectrum, and hence $\ws$ and $\Delta N$, unaffected. Accretion is also negligible in most cases, but becomes non-negligible when the PBH mass spectrum is anomalously broad: in such cases the stasis epoch is abridged and the effective $\Delta N_\mathrm{eff}$ is reduced. The benchmark of Fig.~6 of~\cite{Dienes:2022zgd} used above does not have an anomalously broad spectrum, so accretion corrections are expected to be negligible and the validation stands.

In the more extreme case in which accretion is strong enough to distort the spectrum significantly, $\Omr$ ceases to be strictly constant and the system is no longer in stasis by definition. This caveat is therefore self-limiting: where stasis holds, the template applies; where accretion is large enough to spoil the template, it is also large enough to spoil the underlying stasis attractor, and the question of comparing to a stasis prediction does not arise.

\section{Degeneracies and discriminators}
\label{sec:detectability}

A stasis epoch is not the only way to distort the inflationary gravitational wave background. Early matter domination (EMD), kination, and early dark energy (EDE) all modify the IGWB, and a naive inspection of a power-law slope $\alpha$ between two break frequencies cannot immediately distinguish among them. In this section we characterize which alternatives are genuinely degenerate with stasis in the GW spectrum at the level of the single-epoch template, and identify three independent handles that break the degeneracy.

\subsection{The generic single-fluid result}
\label{sec:degeneracy:generic}

\begin{figure}
\centering
\includegraphics[width=0.9\linewidth]{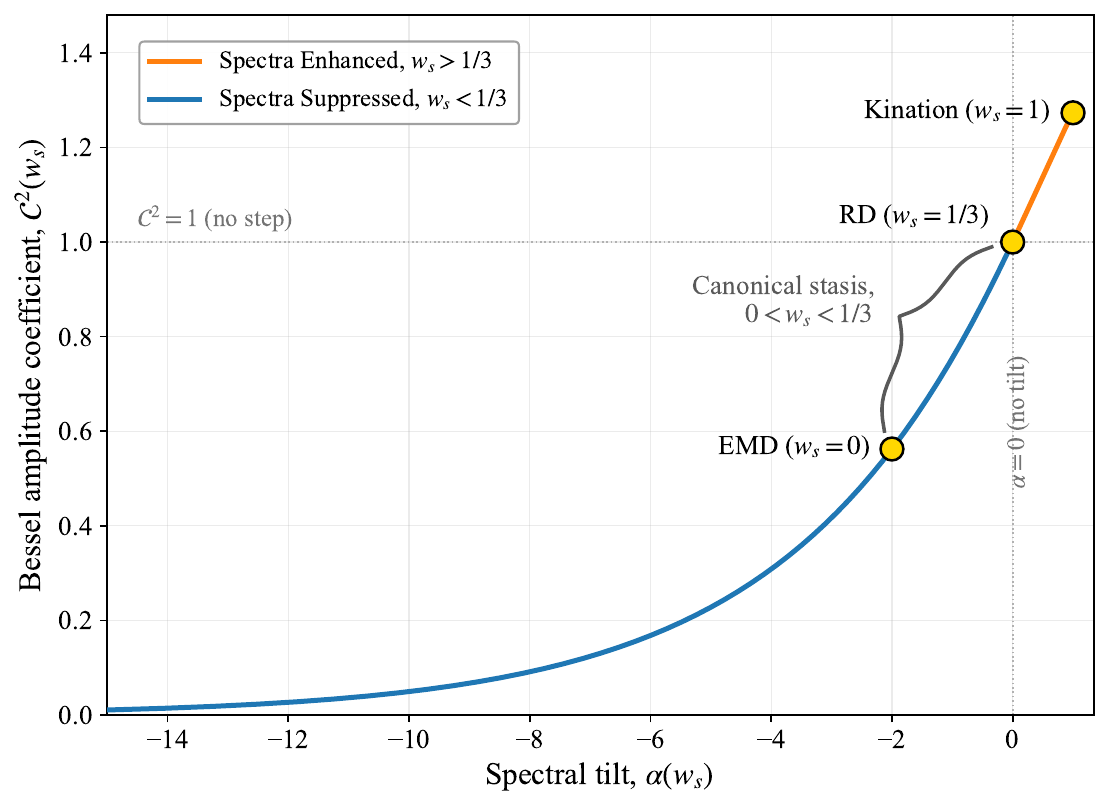}
\caption{Consistency curve $\Ccal^2 = \Ccal^2(\alpha)$~\eqref{eq:consistency_curve}, with the positions of kination, standard radiation domination (RD), and early matter domination (EMD) highlighted using the values from Table~\ref{tab:benchmarks_degeneracy}. As $w_s$ approaches -1/3, the Bessel amplitude coefficient approaches 0, and the spectral tilt diverges to $-\infty$ corresponding to a complete suppression of the spectrum. On the opposite end, the curve terminates at $w_s = 1$, the causal limit of the equation of state.}
\label{fig:consisency_curve}
\end{figure}

Consider any era during which the universe is dominated by a fluid with constant equation of state $w$. The tensor mode equation is
\begin{equation}
h_k'' + 2\mathcal{H} h_k' + k^2 h_k = 0,
\end{equation}
and because the background is an exact power law $a \propto \tau^{2/(1+3w)}$, the solution is an exact Bessel function, identical to the stasis calculation of Sec.~\ref{sec:transfer}. The resulting IGWB template is therefore characterized by three quantities that are all functions of the single number $w$:
\begin{align}
\alpha(w) &= \frac{2(3w-1)}{1+3w}, \label{eq:alpha_generic}\\
\nu(w) &= \frac{3(1-w)}{2(1+3w)}, \label{eq:nu_generic}\\
\Ccal^2(w) &= \left[\frac{2^{\nu+1/2}\,\Gamma(\nu+1)}{\sqrt{\pi}\,\beta^\beta}\right]^2, \quad \beta\equiv\frac{2}{1+3w}. \label{eq:C2_generic}
\end{align}
The key observation is that every constant-$w$ era, whether it arises from stasis, EDE, EMD, or any other mechanism, produces a spectral template that lies on the same consistency curve Eq.~\eqref{eq:consistency_curve} derived in Sec.~\ref{sec:consistency}. The curve is a universal Bessel-function relation, not a stasis-specific prediction. Table~\ref{tab:benchmarks_degeneracy} lists the values at key reference points, and Fig.~\ref{fig:consisency_curve} shows the full curve with these benchmarks highlighted.

\begin{table}[h]
\centering
\renewcommand{\arraystretch}{1.35}
\setlength{\tabcolsep}{18pt}
\begin{tabular}{lcccc}
\toprule
Era & $w$ & $\alpha$ & $\nu$ & $\Ccal^2$ \\
\midrule
Matter domination (MD) & $0$ & $-2$ & $3/2$ & $9/16$ \\
Canonical Stasis with $w_s = 1/6$ & $1/6$ & $-2/3$ & $5/6$ & $0.830$ \\
Radiation domination (RD) & $1/3$ & $0$ & $1/2$ & $1$\\
Kination & $1$ & $+1$ & $0$ & $4/\pi$ \\
\bottomrule
\end{tabular}
\caption{The consistency relation~\eqref{eq:consistency_curve} evaluated at reference cosmologies. All constant-$w$ eras lie on the same curve; the amplitude step $\Ccal^2$ is not a free parameter once $\alpha$ is measured.}
\label{tab:benchmarks_degeneracy}
\end{table}

\subsection{Potential degeneracies}
\label{sec:degeneracy:competitors}

\paragraph{Early matter domination (EMD).} A pressureless fluid gives $w=0$, corresponding to $\alpha=-2$ and $\Ccal^2 = 9/16$. This is the MD endpoint of the canonical stasis range. A stasis epoch with $w_s \to 0$ is observationally identical to EMD in the GW spectrum. However, EMD has no free slope parameter, the spectral index is fixed at $\alpha=-2$. Any measured value $\alpha \neq -2$ with $\Ccal^2$ consistent with Eq.~\eqref{eq:consistency_curve} immediately rules out pure EMD.

\paragraph{Kination.} A kinetically dominated era during which $w=1$ gives $\alpha=+1$ and $\Ccal^2=4/\pi$, corresponding to an enhancement of the IGWB rather than a suppression. For canonical stasis with $0 < w_s < 1/3$, the stasis notch lies at $\alpha < 0$ and kination is not a confusion background. However, stasis models with $w_s > 1/3$ would produce $\alpha > 0$, overlapping in slope with kination; in that case the two may be distinguished by their different values of $\Ccal^2$ on the consistency curve, since kination necessarily sits at the fixed point $\Ccal^2 = 4/\pi$ while stasis with $w_s > 1/3$ may predict a different amplitude step.

\paragraph{Early dark energy (EDE).} An EDE epoch with effective equation of state $w_\mathrm{ede}$ satisfying $-1/3 < w_\mathrm{ede} < 1/3$ produces $-\infty < \alpha < 0$, an overlapping range with that of canonical stasis. If $w_\mathrm{ede}$ is approximately constant during this era, the tensor equation of motion is identical to the stasis equation, and the resulting spectrum lies on exactly the same consistency curve of Eq. ~\eqref{eq:consistency_curve}. A constant-$w$ EDE epoch with $w_\mathrm{ede} = w_s$ is therefore observationally indistinguishable from stasis in the GW spectrum at the level of the single-epoch template. This is the principal degeneracy our framework must address.

\paragraph{A generic non-standard era with varying $w(t)$.} If $w$ varies during the non-standard era, the conformal-time dependence of $\mathcal{H}$ is no longer a pure power law and the Bessel solution breaks down. As a consequence, the tensor spectrum in the corresponding frequency band will deviate from a straight line in log-log space. Stasis, by contrast, predicts an exact power law across the entire stasis band, a direct consequence of the attractor holding $w_s$ constant to high precision. Even realistic EDE models, in which $w_\mathrm{ede}(t)$ deviates from a constant at some level, induce spectral curvature that attractor-enforced stasis does not. Resolving this curvature requires high signal-to-noise across the full width $\Delta N$ of the feature, which high-SNR spectroscopy with BBO for $r \gtrsim 10^{-3}$ could in principle achieve.

\subsection{Discriminators between stasis and its mimics}
\label{sec:degeneracy:discriminators}

Since any single constant-$w$ era lies on the same consistency curve, distinguishing stasis from a generic constant-$w$ mimic requires additional observational handles. Here, we identify three.

\subsubsection*{1. The consistency relation as a self-consistency test}

Although the curve of Eq.~\eqref{eq:consistency_curve} is universal, it constitutes a sharp, non-trivial prediction: measuring $\alpha$ from the spectral slope completely fixes $\Ccal^2$. These are two independently measurable quantities from the GW spectrum: the slope in the stasis band from the spectral index between $\fend$ and $\fbeg$ and the amplitude step at $\fend$ from the ratio of the signal just below and just above the lower break. Any model that breaks the constant-$w$ approximation such as a slowly varying EOS, a two-component fluid, or any other modification, will generically violate Eq.~\eqref{eq:consistency_curve}. The self-consistency test is therefore be to first measure $\alpha$ from the slope, then predict $\Ccal^2$, then compare this prediction with the measured amplitude step.

Any discrepancy in this comparison falsifies the constant-$w$ hypothesis. The stasis attractor enforces this consistency, so a confirmed consistency relation provides evidence for a mechanism with strong attractor behavior. We quantify how precisely BBO and DECIGO can perform this test in Sec.~\ref{sec:fisher}.

\subsubsection*{2. The multi-epoch comb as a combinatorial test}

The strongest discriminator arises from the multi-epoch template. If stasis recurs in multiple distinct epochs with parameters $(\ws^{(i)}, \Delta N^{(i)})$, the GW spectrum exhibits a comb of notches, each satisfying its own consistency relation $\Ccal^2_i = \Ccal^2(\alpha_i)$, with inter-plateau amplitude ratios
\begin{equation}
\frac{\Omega_{\mathrm{GW}}^{(i+1)\,\text{plateau}}}{\Omega_{\mathrm{GW}}^{(i)\,\text{plateau}}} = \left(\frac{\fbeg^{(i)}}{\fend^{(i)}}\right)^{\alpha_i}.
\end{equation}
A coincidental sequence of $N$ independent non-standard eras would require each amplitude ratio to match its independently predicted $\Ccal^2(\alpha_i)$ and all inter-plateau ratios to be consistent with the accumulated tilt products. The probability of this occurring by chance decreases rapidly with $N$. Since EDE typically involves a single scalar field and produces a single spectral feature, this comb structure has no EDE analogue: a confirmed comb with $N \geq 2$ notches, each passing the consistency test, provides very strong evidence for stasis specifically.  A coincidental sequence of $N$ independent non-standard eras would require each notch to
independently satisfy its own consistency relation $\mathcal{C}^2_i = \mathcal{C}^2(\alpha_i)$ and the
inter-plateau ratios to match the accumulated tilt predictions. The probability of this occurring by accident
decreases exponentially with $N$, making a confirmed comb with $N \geq 2$ notches essentially impossible to
attribute to unrelated non-standard epochs. We discuss this in more detail in our forthcoming work \cite{BarenboimBurns:stasis2}.

\subsubsection*{3. Independent constraints on $w_s$ and $\Delta N$ from particle physics}

In concrete realizations of stasis such as the PBH tower or the KK graviton tower neither $w_s$ nor $\Delta N$ is a free parameter: both are predicted from the underlying particle physics. The attractor value $w_s = \Omr/3$ is fixed in terms of the tower mass spectrum $\delta$ and evaporation parameter $\alpha_\mathrm{PBH}$, and $\Delta N$ follows from the same parameters. EDE, by contrast, has no analogous prediction for $\Delta N$, which is a free input. If the $w_s$ inferred from the GW slope $\alpha$ matches the value predicted from tower parameters constrained by independent observations such as structure formation or PBH abundance limits, and if $\Delta N$ can be independently bounded from the CMB, the degeneracy with generic EDE is broken by multi-messenger consistency rather than by the GW template alone.

\subsection{Relation to prior work}
\label{sec:degeneracy:prior_work}

The functional form $\alpha(\ws) = 2(3\ws-1)/(1+3\ws)$ relating the spectral tilt to the equation of state, together with the Bessel coefficient $\Ccal^2(\nu)$ in Eq.~\eqref{eq:Ccal}, is not new in isolation. For specific non-standard eras, both quantities have appeared in the literature for many years. Giovannini, et al. computed the inflationary GW spectrum for a stiff post-inflationary epoch with $\ws = 1$ and obtained the corresponding tilt and amplitude~\cite{Giovannini:1998bp,Giovannini:1999bh}. The Boyle--Steinhardt analysis~\cite{Boyle:2005se} extended this systematically to arbitrary constant $w$, presenting $\Ccal^2(\nu)$ in essentially the form used here and tabulating the matter-dominated and radiation-dominated benchmarks. Subsequent work on EMD scenarios established $\alpha(0) = -2$ as the canonical signature of an intermediate matter-dominated era~\cite{Seto:2003kc,Boyle:2005se}, and the kination tilt $\alpha(1) = +1$ has been studied extensively as a probe of stiff post-inflationary phases~\cite{Co:2021lkc}.

What is new here is threefold. First we present the $(\alpha, \Ccal^2)$ pairing as an observational discriminator rather than a purely calculational tool. Second, we extend the consistency relation to multi-component stasis cosmologies in which $\ws$ is set by attractor abundances rather than by a single dominant fluid. Finally, we show its preservation across multi-epoch combs, where each break independently tests the same one-parameter relation.

\section{Fisher forecast for BBO and DECIGO}
\label{sec:fisher}

Having established the consistency relation as the key discriminator in Sec.~\ref{sec:detectability}, we now quantify how precisely next-generation GW detectors can measure the spectral tilt, $\alpha(w_s)$ and the bessel amplitude coefficient, $\Ccal^2$ jointly, and therefore how sensitively they can test whether a measured spectral feature lies on the curve~\eqref{eq:consistency_curve}.

\subsection{Setup and signal model}
\label{sec:fisher:setup}

In order to quantify the detectability of the stasis models described above by the future detectors BBO \cite{Corbin:2005ny} and DECIGO \cite{Kawamura:2011zz}, we perform a Fisher analysis based on our piecewise spectral template introduced in Section \ref{sec:template}. The template has five parameters, $\alpha$, $\Ccal^2$, $\fend$, $\fbeg$, and $\Omega_\mathrm{RD}$, of which we estimate $\alpha$ and $\Ccal^2$ and hold the remaining three fixed. Since $\fend$, $\fbeg$, and $\Omega_\mathrm{RD}$ set the location and overall amplitude of the stasis band rather than the shape that encodes the consistency relation, choosing their values isolates the two parameters of interest and allows us to determine a best-case forecast for how well $\alpha$ and $\Ccal^2$ can be jointly measured. We choose $\fend$ and $\fbeg$ so that the stasis band falls within the most sensitive frequency range of BBO and DECIGO, making the resulting uncertainties optimistic by construction. The signal model is
\begin{equation}
    \Omega_\mathrm{GW}(f) = \Omega_\mathrm{RD} \times
    \begin{cases}
        1 & f < \fend \\
        \Ccal^2 \left(f/\fend\right)^{\alpha} & \fend \leq f < \fbeg \\
        \left(\fbeg/\fend\right)^{\alpha} & f \geq \fbeg
    \end{cases}
    \label{eq:fisher_signal}
\end{equation}
where $\Omega_\mathrm{RD}$ is the post-stasis radiation-dominated plateau amplitude.
 
We model the instrumental sensitivity of the two detectors with the effective noise energy density per frequency interval in log space, $h^2\Omega_n(f) = (4\pi^2/3)\, f^3 S_n(f)/H_{100}^2$, where $S_n(f)$ is the one-sided strain noise power spectral density (PSD) and $H_{100} = 100\,\mathrm{km\,s^{-1}\,Mpc^{-1}}$. For both instruments we adopt a broken-power-law form for the strain noise PSD, $S_n(f) = S_0\,[(f_0/f)^4 + 1 + (f/f_0)^2]$, with $(S_0, f_0) = (1.44\times10^{-54}\,\mathrm{Hz^{-1}},\,0.30\,\mathrm{Hz})$ for BBO and $(S_0, f_0) = (2.3\times10^{-52}\,\mathrm{Hz^{-1}},\,0.35\,\mathrm{Hz})$ for DECIGO. The two terms scaling as $f^{-4}$ and $f^{2}$ reproduce the low-frequency acceleration-noise term and the high-frequency shot-noise term, respectively, while the constant term fixes the floor; the resulting energy-density floors are $O(10^{-19})$ near $0.2\,\mathrm{Hz}$ for BBO and $O(10^{-17})$ near $0.23\,\mathrm{Hz}$ for DECIGO, calibrated to the published single cross-correlation-pair noise floors of Refs.~\cite{Corbin:2005ny,Kawamura:2011zz}. We combine the two instruments by using the BBO curve for $f \leq 1\,\mathrm{Hz}$ and the DECIGO curve for $f > 1\,\mathrm{Hz}$, and we assume a total integrated observation time $T_\mathrm{obs} = 4\,\mathrm{yr}$. 

\subsection{The Fisher matrix}
\label{sec:fisher:matrix}

To determine the joint sensitivity of BBO and DECIGO to our stasis models, we estimate $\theta = (\alpha,\, \ln\Ccal^2)$, using $\ln\Ccal^2$ so that the derivative with respect to $\ln\Ccal^2$ is scale-free. For a stochastic background measured by cross-correlation, the measured energy density in each frequency bin is Gaussian-distributed and the variance is set by the detector noise and observation time. The $2\times2$ Fisher matrix is
\begin{equation}
    F_{ij} = T_\mathrm{obs} \int \left(\frac{\Omega_\mathrm{GW}(f)}{\Omega_n(f)}\right)^2
    \frac{\partial \ln \Omega_\mathrm{GW}}{\partial \theta_i}
    \frac{\partial \ln \Omega_\mathrm{GW}}{\partial \theta_j}
    \, d\ln f ,
\end{equation}
where $\Omega_n(f) \equiv h^2\Omega_n(f)$ is the combined noise energy density introduced above and the integral runs over the detector band. Because the template is a piecewise power law, the logarithmic derivatives are analytic,
\begin{align}
    \frac{\partial \ln \Omega_\mathrm{GW}}{\partial \alpha} &=
    \begin{cases}
        \ln(f/\fend) & \fend \leq f < \fbeg \\
        \ln(\fbeg/\fend) & f \geq \fbeg \\
        0 & f < \fend
    \end{cases} \\
    \frac{\partial \ln \Omega_\mathrm{GW}}{\partial \ln \Ccal^2} &=
    \begin{cases}
        1 & \fend \leq f < \fbeg \\
        0 & \text{otherwise} .
    \end{cases}
\end{align}
Thus the amplitude is constrained only in the stasis band $\fend \leq f < \fbeg$, while the slope is constrained both in the band and in the pre-stasis plateau. Inside the band the $\alpha$-derivative $\ln(f/\fend)$ varies with frequency, breaking the degeneracy with $\Ccal^2$, above $\fbeg$ the $\Ccal^2$-derivative vanishes and the $\alpha$-derivative is the constant $\ln(\fbeg/\fend)$, so the plateau height $\alpha\ln(\fbeg/\fend)$ constrains $\alpha$ alone and carries no information on $\Ccal^2$. The off-diagonal element $F_{\alpha,\ln\Ccal^2} = T_\mathrm{obs}\int (\Omega_\mathrm{GW}/\Omega_n)^2 \ln(f/\fend)\, d\ln f$ is always positive over the band, encoding the physical degeneracy between slope and amplitude. The covariance matrix is $\Sigma = F^{-1}$, and the marginalized $1\sigma$ uncertainties are $\sigma_\alpha = \sqrt{\Sigma_{00}}$ and $\sigma_{\Ccal^2} = \Ccal^2\sqrt{\Sigma_{11}}$, the latter converting from $\ln\Ccal^2$ back to $\Ccal^2$.

\subsection{PBH tower benchmarks}
\label{sec:fisher:benchmarks}
For concreteness, we analyze four benchmark points in the PBH stasis tower model, labeled by the spectral index of the PBH mass distribution, $\alpha_\mathrm{PBH} \in \{-5/4,\,-3/2,\,-7/4,\,-2\}$. Each characterizes a constant-$w$ stasis era with equation-of-state parameter $w_s$ and duration $\Delta N$, which fix the template slope $\alpha = \alpha(w_s)$ and Bessel amplitude $\Ccal^2 = \Ccal^2(w_s)$ on the consistency curve. We take $\fend$ at $T_\mathrm{end} = 10^7\,\mathrm{GeV}$, which places $\fend \simeq 0.265\,\mathrm{Hz}$ squarely in the BBO band. Table~\ref{tab:benchmarks} collects the parameters and the resulting uncertainties at the fiducial tensor-to-scalar ratio $r = 0.01$. The detection threshold $r_\mathrm{min}$ is defined by $\mathrm{SNR}_\mathrm{feature} = 1$, where $\mathrm{SNR}_\mathrm{feature}^2 = T_\mathrm{obs}\int [\,\Omega_\mathrm{RD}\,|1 - \mathcal{M}(f)|/\Omega_n(f)\,]^2\, d\ln f$ measures the integrated significance of the departure of the band shape, $\mathcal{M}(f)$, from the flat radiation-dominated reference. In the most challenging to detect of the four scenarios in which $\alpha_\mathrm{PBH} = -5/4$, we find $\sigma_\alpha \simeq 6\times10^{-5}$ and $\sigma_{\Ccal^2} \simeq 6\times10^{-6}$ at $r = 0.01$, the loosest of the four benchmarks and hence the most conservative best-case bound.
 
\begin{table}
\renewcommand{\arraystretch}{1.35}
\setlength{\tabcolsep}{8pt}
\centering
\begin{tabular}{lcccccc}
\hline\hline
$\alpha_\mathrm{PBH}$ &  $\alpha$ & $\Ccal^2$ & $r_\mathrm{min}$ & $\sigma_\alpha$ & $\sigma_{\Ccal^2}$ & $\sigma_\perp$ \\
\hline
$-5/4$ & $-1.543$ & $0.644$ & $4.8\times10^{-8}$ & $6.1\times10^{-5}$ & $6.2\times10^{-6}$ & $1.6\times10^{-5}$ \\
$-3/2$  & $-1.142$ & $0.724$ & $5.9\times10^{-8}$ & $4.8\times10^{-5}$ & $5.9\times10^{-6}$ & $1.5\times10^{-5}$ \\
$-7/4$  & $-0.799$ & $0.800$ & $7.9\times10^{-8}$ & $3.9\times10^{-5}$ & $5.7\times10^{-6}$ & $1.3\times10^{-5}$ \\
$-2$    & $-0.500$ & $0.871$ & $1.2\times10^{-7}$ & $3.3\times10^{-5}$ & $5.5\times10^{-6}$ & $1.2\times10^{-5}$ \\
\hline\hline
\end{tabular}
\caption{PBH tower benchmarks. The label $\alpha_\mathrm{PBH}$ is the spectral index of the PBH mass distribution and should not be confused with the template slope $\alpha = \alpha(w_s)$ that the Fisher analysis measures. Uncertainties $\sigma_\alpha$, $\sigma_{\Ccal^2}$, and the perpendicular-to-curve resolution $\sigma_\perp$ are quoted at $r = 0.01$ for $T_\mathrm{obs} = 4\,\mathrm{yr}$ and $T_\mathrm{end} = 10^7\,\mathrm{GeV}$; all three scale as $r^{-1}$. The detection threshold $r_\mathrm{min}$ is set by $\mathrm{SNR}_\mathrm{feature} = 1$. Please note that the values of $r_\mathrm{min}$ quoted here include the full $T_\mathrm{obs} = 4\,\mathrm{yr}$ integration in $\mathrm{SNR}_\mathrm{feature}$. Note that these are optimistic bounds, as foreground contamination would raise the effective detection threshold above the idealized $\mathrm{SNR}_\mathrm{feature}$ = 1 assumed here.}
\label{tab:benchmarks}
\end{table}

\subsection{Results: ellipses on the consistency curve}
\label{sec:fisher:results}
\begin{figure}
\centering
\includegraphics[width=0.9\linewidth]{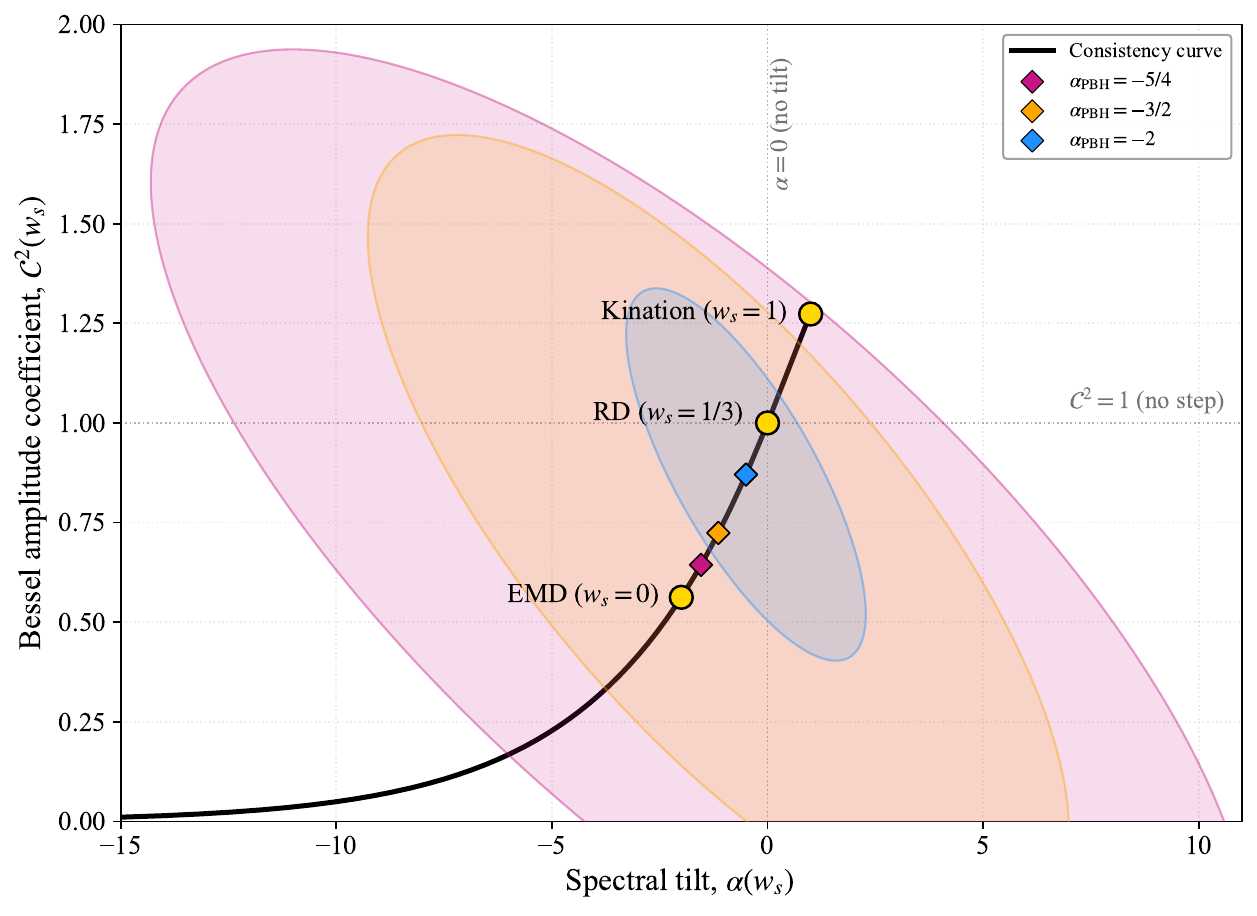}
\caption{Full $(\alpha,\Ccal^2)$ consistency curve with $1\sigma$ Fisher ellipses for
three PBH benchmarks, shown at the threshold amplitude $r = r_\mathrm{min}$, the largest
ellipses consistent with a detection; BBO+DECIGO, $T_\mathrm{obs} = 4\,\mathrm{yr}$,
$T_\mathrm{end} = 10^7\,\mathrm{GeV}$. At threshold $\sigma_\perp$ is
comparable to the curve's range in $\Ccal^2$, so the ellipses swell beyond it. The $\alpha_\mathrm{PBH} = -7/4$ benchmark is omitted as its ellipse was visually difficult
to distinguish from that of $\alpha_\mathrm{PBH} = -2$.}
\label{fig:fisher_a}
\end{figure}

Figure~\ref{fig:fisher_a} shows the full consistency curve in the $(\alpha, \Ccal^2)$
plane together with $1\sigma$ Fisher ellipses for three benchmarks. We plot them at the threshold amplitude $r = r_\mathrm{min}$ at which each feature reaches $\mathrm{SNR} = 1$ so that the ellipses are visible on the scale of the full curve. These, therefore, are the largest ellipses consistent with a detection. The ellipses are strongly anisotropic: the slope--amplitude degeneracy encoded in the positive off-diagonal
$F_{\alpha,\ln\Ccal^2}$ produces a negative correlation between $\alpha$ and $\Ccal^2$. The long axis is misaligned with the local tangent of the consistency curve, so what matters for the consistency test is the extent measured perpendicular to it. We therefore adopt as the key figure of merit the perpendicular width
\begin{equation}
    \sigma_\perp \equiv \sqrt{\hat{n}^\top \Sigma \, \hat{n}},
    \qquad \hat{n} = \frac{(-\partial_\alpha \Ccal^2,\; 1)}
    {\sqrt{1 + (\partial_\alpha \Ccal^2)^2}} ,
\end{equation}
where $\hat{n}$ is the unit normal to the curve and $\Sigma$ is expressed in $(\alpha, \Ccal^2)$ coordinates. As shown in Fig.~\ref{fig:fisher_a}, for the $r = r_\mathrm{min}$ case, the ellipses swell well beyond the curve, with
$\sigma_\perp \simeq 1$--$3$, comparable to or larger than the curve's $\Ccal^2$ range,
$\Delta\Ccal^2 \simeq 0.7$ between the matter-dominated and kination endpoints, so the
test has little power at the detection threshold. Because $\sigma_\perp \propto 1/r$,
however, it shrinks rapidly above threshold: at $r = 0.01$ (Sec.~\ref{sec:fisher:zoom})
we find $\sigma_\perp \simeq 1.2$--$1.6\times10^{-5}$, four to five orders of magnitude below $\Delta\Ccal^2$. A constant-$w$ era lands on the curve by construction. We find that for any amplitude comfortably above threshold $\sigma_\perp$ is minuscule compared to the off-curve displacements such models produce, so the consistency test retains strong discriminating power across the detectable range of $r$.
 
\subsection{Zoom: resolving individual benchmarks}
\label{sec:fisher:zoom}

\begin{figure}
\centering
\includegraphics[width=0.9\linewidth]{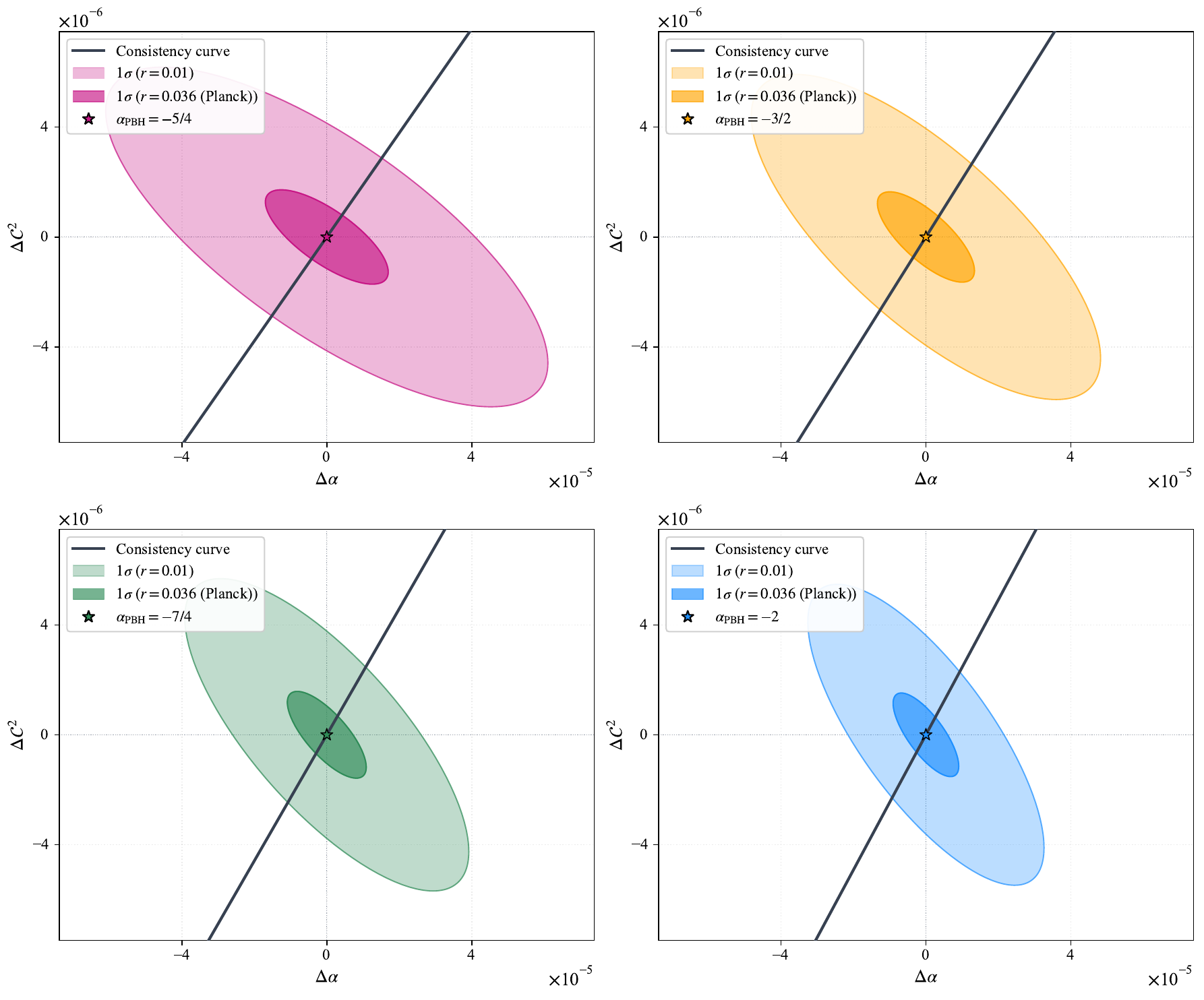}
\caption{Zoom on the four benchmarks at $r = 0.01$ (light) and $r = 0.036$ (dark). The
ellipse shrinks as the stasis equation of state approaches radiation domination: a
shallower tilt and a larger amplitude coefficient $\Ccal^2$ both keep signal-to-noise
high across the band, which the steeper benchmarks lose at high frequency.}
\label{fig:fisher_b}
\end{figure}

Figure~\ref{fig:fisher_b} zooms in on the individual benchmarks at $r = 0.01$ and $r = 0.036$. The constraint tightens as the stasis equation of state approaches radiation domination: at $r = 0.01$, $\sigma_\alpha$ falls from $6.1\times10^{-5}$ for the $\alpha_\mathrm{PBH} = -5/4$ benchmark to $3.3\times10^{-5}$ for $\alpha_\mathrm{PBH} = -2$. Two effects, both controlled by $w_s$, set the trend and act in the same direction. First, a more negative tilt suppresses $\Omega_\mathrm{GW}$ toward higher frequencies,  where $\partial_\alpha \ln\Omega_\mathrm{GW} = \ln(f/\fend)$ is largest, so the steeper benchmarks lose Fisher information at high frequency. Second, the amplitude coefficient $\Ccal^2$ is smaller for those same benchmarks ($\Ccal^2 \simeq 0.64$ at $\alpha_\mathrm{PBH} = -5/4$ versus $0.87$ at $-2$), lowering the overall signal, so benchmarks nearer radiation domination are measured most tightly. All uncertainties scale inversely with the tensor amplitude,
\begin{equation}
    \sigma_\alpha \propto r^{-1}, \qquad \sigma_{\Ccal^2} \propto r^{-1}, \qquad
    \sigma_\perp \propto r^{-1} ,
\end{equation}
since the Fisher matrix scales as $F \propto \Omega_\mathrm{GW}^2 \propto \Omega_\mathrm{RD}^2 \propto r^2$ and hence $\Sigma = F^{-1} \propto r^{-2}$; halving $r$ therefore doubles the ellipse axes and quadruples its area.

\subsection{Detection threshold and observational prospects}
\label{sec:fisher:prospects}

There are three conclusions that can be drawn from this forecast. First, at the fiducial $r = 0.01$ with $\Delta N \geq 6$, BBO+DECIGO resolve the consistency relation at very high significance. $\sigma_\perp \simeq 1.5\times10^{-5}$ is four orders of magnitude below the range of $\Ccal^2$, so any constant-$w$ era is pinned to the curve and any off-curve deviation can be detected with high precision. Second, the test does not require the full Planck-allowed amplitude, $r = 0.036$ and the spectral feature is detectable for all four benchmarks down to very small values of r. Third, detectability of the feature does not by itself guarantee a sharp consistency test: because $\sigma_\perp \propto r^{-1}$, the perpendicular resolution degrades as $r$ approaches $r_\mathrm{min}$, and at threshold ($\mathrm{SNR}_\mathrm{feature} = 1$) it grows to $\sigma_\perp \sim 1$--$3$, comparable to or larger than the curve's extent, so the ellipse is larger than the vertical span of the curve and the discriminating power is lost. The test remains sharp down to $r \sim 10^{-6}$, which still spans the upper several decades of the detectable range and includes the full Planck-allowed window. Cases with shallow slopes that are close in parameter space to radiation domination are easiest to constrain, while steep slopes in which $\alpha$ approaches $-2$ and $w_s$ approaches 0 are harder, requiring a larger $r$ to reach the same perpendicular resolution.

\section{Discussion and outlook}
\label{sec:discussion}

In this study we have shown that a generic cosmological stasis epoch leaves a detectable imprint on the inflationary gravitational wave background. Because stasis requires a constant equation of state $\ws$, the tensor mode equation reduces to an exact Bessel equation, producing closed forms of both the spectral distortion $\alpha(\ws)$ and the Bessel amplitude coefficient $\Ccal^2(w_s)$. Our piecewise template has only two physical inputs, the equation of state $\ws$ and the duration $\Delta N_\mathrm{stasis}$, and applies uniformly across every microphysical realization of stasis, from decaying towers and primordial black holes to dynamical scalars, thermal annihilation, and purely gravitational interactions. The central result is the consistency relation of Eq.~\eqref{eq:consistency_curve}. By eliminating $\ws$ between the measured stasis-band slope $\alpha$ and the amplitude step $\Ccal^2$ we show that every constant-$w$ era is a point on a single curve, turning the transfer function into a falsifiable prediction. 

In Section \ref{sec:validation}, we validated the stasis-band slope against the numerical PBH-induced spectrum of~\cite{Dienes:2022zgd} with no free parameters. In Section \ref{sec:fisher} we showed that BBO and DECIGO can resolve the perpendicular displacement from the consistency curve to $\sigma_\perp \simeq 1.5\times10^{-5}$ at $r = 0.01$, four orders of magnitude below the curve's range in $\Ccal^2$ meaning that any off-curve deviation is detectable across nearly the full Planck-allowed window. 

The present template models both spectral breaks as sharp transitions and assumes a standard radiation-dominated pre-stasis cosmology. Relaxing these assumptions, together with a quantitative test of the plateau-ratio prediction in realizations whose pre-history is genuinely radiation dominated, is left to future work, as is the treatment of perturbations during stasis~\cite{Dienes:2025tox}. Some of these directions are taken up in two companion papers that will follow shortly. In the second paper in this series~\cite{BarenboimBurns:stasis2} we extend the single-epoch template to multi-epoch stasis combs. In the third paper in the series~\cite{BarenboimBurns:stasis3} we map the $(\ws, \Delta N_\mathrm{stasis})$ plane onto the sensitivities of LISA, DECIGO, BBO, the Einstein Telescope, Cosmic Explorer, and pulsar timing arrays. In addition, we identify which microphysical realizations land in which detectors' bands, incorporate the $g_*$ fine structure, and model the smooth end-of-stasis transition and its associated observables. These results establish the inflationary gravitational wave background as a direct and precise probe of pre-BBN stasis epochs at energy scales far higher than the ones accessible by any other cosmological observable.

\section*{Acknowledgments}

We would like to thank Tim Tait for the valuable feedback on our work. GB is supported by the Spanish grant  PID2023-151418NB-I00 funded by MCIU/AEI/10.13039/501100011033. AKB acknowledges support from the ``Unit of Excellence Maria de Maeztu 2020-2023'' award to the ICC-UB CEX2019-000918-M and grant PID2022-136224NB-C21 funded by MCIN / MINECO / MCOC. AKB and GB are supported by the European Union’s Horizon 2020 research and innovation program under the Marie Skłodowska-Curie grant agreement No 860881-HIDDeN, and Horizon Europe research and innovation program under the Marie Skłodowska-Curie Staff Exchange grant agreement No 101086085 – ASYMMETRY. 

\bibliographystyle{apsrev4-2}
\bibliography{biblio}

\end{document}